\documentclass[12pt]{article}
\usepackage{fancyhdr}

\usepackage{mathrsfs}
\usepackage[T1]{fontenc}
\usepackage{setspace}
\usepackage{amsfonts}
\usepackage{amssymb}
\usepackage{amsmath}
\usepackage{epsfig}
\usepackage{latexsym}
\usepackage{color}
\usepackage{graphicx}
\usepackage{nicefrac}
\usepackage[latin1]{inputenc}
\usepackage{slashed}
\usepackage{multirow}
\usepackage{comment}
\usepackage{soul}
\usepackage{hyperref}
\usepackage{cite}
\usepackage{slashed}
\usepackage{simpler-wick}
\usepackage{array}
\usepackage{tabu}
\usepackage{tcolorbox}

\usepackage[titletoc]{appendix}

\def\hybrid{\topmargin -20pt    \oddsidemargin 0pt
        \headheight 0pt \headsep 0pt
        \textwidth 6.25in       
        \textheight 9.25in       
        \marginparwidth .875in
        \parskip 5pt plus 1pt   \jot = 1.5ex}

\hybrid

\def\baselinestretch{1.2}

\catcode`\@=11

\def\marginnote#1{}
\newcount\hour
\newcount\minute
\newtoks\amorpm
\hour=\time\divide\hour by60
\minute=\time{\multiply\hour by60 \global\advance\minute by-\hour}
\edef\standardtime{{\ifnum\hour<12 \global\amorpm={am}%
        \else\global\amorpm={pm}\advance\hour by-12 \fi
        \ifnum\hour=0 \hour=12 \fi
        \number\hour:\ifnum\minute<10 0\fi\number\minute\the\amorpm}}
\edef\militarytime{\number\hour:\ifnum\minute<10 0\fi\number\minute}

\def\draftlabel#1{{\@bsphack\if@filesw {\let\thepage\relax
   \xdef\@gtempa{\write\@auxout{\string
      \newlabel{#1}{{\@currentlabel}{\thepage}}}}}\@gtempa
   \if@nobreak \ifvmode\nobreak\fi\fi\fi\@esphack}
        \gdef\@eqnlabel{#1}}
\def\@eqnlabel{}
\def\@vacuum{}
\def\draftmarginnote#1{\marginpar{\raggedright\scriptsize\tt#1}}

\def\draft{\oddsidemargin -.5truein
        \def\@oddfoot{\sl preliminary draft \hfil
        \rm\thepage\hfil\sl\today\quad\militarytime}
        \let\@evenfoot\@oddfoot \overfullrule 3pt
        \let\label=\draftlabel
        \let\marginnote=\draftmarginnote
   \def\@eqnnum{(\theequation)\rlap{\kern\marginparsep\tt\@eqnlabel}%
\global\let\@eqnlabel\@vacuum}  }

\def\preprint{\twocolumn\sloppy\flushbottom\parindent 2em
        \leftmargini 2em\leftmarginv .5em\leftmarginvi .5em
        \oddsidemargin -.5in    \evensidemargin -.5in
        \columnsep .4in \footheight 0pt
        \textwidth 10.in        \topmargin  -.4in
        \headheight 12pt \topskip .4in
        \textheight 6.9in \footskip 0pt
        \def\@oddhead{\thepage\hfil\addtocounter{page}{1}\thepage}
        \let\@evenhead\@oddhead \def\@oddfoot{} \def\@evenfoot{} }

\def\numberbysection{\@addtoreset{equation}{section}
        \def\theequation{\thesection.\arabic{equation}}}

\def\underline#1{\relax\ifmmode\@@underline#1\else
        $\@@underline{\hbox{#1}}$\relax\fi}

\def\titlepage{\@restonecolfalse\if@twocolumn\@restonecoltrue\onecolumn
     \else \newpage \fi \thispagestyle{empty}\c@page\z@
        \def\thefootnote{\fnsymbol{footnote}} }

\def\endtitlepage{\if@restonecol\twocolumn \else \newpage \fi
        \def\thefootnote{\arabic{footnote}}
        \setcounter{footnote}{0}}  

\catcode`@=12
\relax

\def\figcap{\section*{Figure Captions\markboth
        {FIGURECAPTIONS}{FIGURECAPTIONS}}\list
        {Figure \arabic{enumi}:\hfill}{\settowidth\labelwidth{Figure
999:}
        \leftmargin\labelwidth
        \advance\leftmargin\labelsep\usecounter{enumi}}}
 \relax
\def\tablecap{\section*{Table Captions\markboth
        {TABLECAPTIONS}{TABLECAPTIONS}}\list
        {Table \arabic{enumi}:\hfill}{\settowidth\labelwidth{Table
999:}
        \leftmargin\labelwidth
        \advance\leftmargin\labelsep\usecounter{enumi}}}
 \relax
\def\reflist{\section*{References\markboth
        {REFLIST}{REFLIST}}\list
        {[\arabic{enumi}]\hfill}{\settowidth\labelwidth{[999]}
        \leftmargin\labelwidth
        \advance\leftmargin\labelsep\usecounter{enumi}}}
 \relax
\makeatletter
\newcounter{pubctr}
\def\publist{\@ifnextchar[{\@publist}{\@@publist}}
\def\@publist[#1]{\list
        {[\arabic{pubctr}]\hfill}{\settowidth\labelwidth{[999]}
        \leftmargin\labelwidth
        \advance\leftmargin\labelsep
        \@nmbrlisttrue\def\@listctr{pubctr}
        \setcounter{pubctr}{#1}\addtocounter{pubctr}{-1}}}
\def\@@publist{\list
        {[\arabic{pubctr}]\hfill}{\settowidth\labelwidth{[999]}
        \leftmargin\labelwidth
        \advance\leftmargin\labelsep
        \@nmbrlisttrue\def\@listctr{pubctr}}}
 \relax
\makeatother
\newskip\humongous \humongous=0pt plus 1000pt minus 1000pt

\newif\ifdtup

\relax

\def\be{\begin{equation}}
\def\ee{\end{equation}}
\def\ba{\begin{eqnarray}}
\def\ea{\end{eqnarray}}

\def\del{\partial}

\def\k{\kappa}

\def\a{\alpha}

\def\g{\gamma}
\def\G{\Gamma}
\def\d{\delta}
\def\D{\Delta}
\def\e{\epsilon}

\def\p{\pi}

\def\m{\mu}
\def\n{\nu}

\def\l{\lambda}
\def\L{\Lambda}
\def\s{\sigma}

\def\no{\noindent}

\def\IR{\relax{\rm I\kern-.18em R}}

\def\IR{\relax{\rm I\kern-.18em R}}
\def\IL{\relax{\rm I\kern-.18em L}}

\def\inv{^{\raise.15ex\hbox{${\scriptscriptstyle -}$}\kern-.05em 1}}

\def\cO{{\cal O}}

\def\Tr{{\rm Tr}}

\begin{document}

\renewcommand{\theequation}{\thesection.\arabic{equation}}
\csname @addtoreset\endcsname{equation}{section}

\newcommand{\beq}{\begin{equation}}
\newcommand{\eeq}[1]{\label{#1}\end{equation}}
\newcommand{\ber}{\begin{equation}}
\newcommand{\eer}[1]{\label{#1}\end{equation}}
\newcommand{\eqn}[1]{(\ref{#1})}
\begin{titlepage}
\begin{center}

${}$
\vskip .2 in

{\Large\bf   Yukawa interactions at Large Charge}

\vskip 0.4in

{\bf Oleg Antipin, Jahmall Bersini, Pantelis Panopoulos}
\vskip 0.1in

\vskip 0.1in

 {\em
Rudjer Boskovic Institute, Division of Theoretical Physics, \\
Bijenicka 54, 10000 Zagreb, Croatia}

\vskip 0.1in

{\footnotesize  \texttt { oantipin@irb.hr, jbersini@irb.hr, Pantelis.Panopoulos@irb.hr}}

\today
\vskip .5in
\end{center}

\centerline{\bf Abstract}

\no
We extend the fixed-charge semiclassical method by computing anomalous dimensions of fixed-charge scalar operators in models with Yukawa interactions.
In particular, we discuss the Nambu-Jona-Lasinio-Yukawa theory as well as an asymptotically safe gauge-Yukawa model in four dimensions. 
In the weakly coupled regime, we cross-check our results to the respective maximum known orders of perturbation theory in these models and predict higher order terms for future comparisons with other computational methods. In the strongly coupled  regime, we match our results to the predictions of the large-charge effective field theory which can be compared to future Monte Carlo and lattice studies.

{\footnotesize  \it Preprint: RBI-ThPhys-2022-30}

\vskip .4in
\noindent
\end{titlepage}
\vfill
\eject

\newpage

\tableofcontents

\noindent

\def\baselinestretch{1.2}
\baselineskip 20 pt
\noindent



\setcounter{equation}{0}

\renewcommand{\theequation}{\thesection.\arabic{equation}}

\section{Introduction}

Quantum Field Theory (QFT) is the main language for describing quantum phenomena below atomic scales. In weak coupling regions, perturbation theory is surprisingly successful modelling the systems as a small deformations from the free field theory case that can be solved exactly. 
In strongly coupled regime perturbation theory breaks down and  we should consider different routes of approach. Nowadays,  several  methods have been  developed among which are electric-magnetic duality relating different energy regions of the system \cite{Seiberg:1994rs,Seiberg:1994pq, Seiberg:2016gmd} and the AdS/CFT  correspondence \cite{Maldacena:1997re,DHoker:2002nbb, Aharony:2008ug,Aharony:1999ti} where the weakly coupled string theory in the bulk is mapped to a strongly coupled theory on the boundary of space. Also, using   localization principle we can calculate the non-perturbative partition function of a supersymmetric QFT and correlation functions of  supersymmetric  operators in various dimensions \cite{Pestun:2016zxk}. 

Another appealing method in this line of non-perturbative approach  is the seminal work of \cite{Hellerman:2015nra} where the scaling dimensions of operators with large $U(1)$ charge in 3$D$ conformal field theories (CFTs) were computed using an effective field theory approach as an expansion in inverse powers of the conserved charge. 
Later, numerous applications of the large charge expansion have taken place ranging from scalar field theories \cite{Badel:2019oxl, Antipin:2020abu, Antipin:2021jiw, Giombi:2022gjj, Giombi:2020enj, Watanabe:2022htq, Dondi:2022wli, Alvarez-Gaume:2016vff, Orlando:2019skh, Jack:2021lja, Jack:2020wvs, Cuomo:2021ygt, Arias-Tamargo:2019xld,Antipin:2022dsm,Bednyakov:2022guj} to non-relativistic CFT \cite{Hellerman:2021qzz, Pellizzani:2021hzx} and supersymmetric theories \cite{Hellerman:2017sur,Hellerman:2020sqj,Sharon:2020mjs,Beccaria:2018xxl,Hellerman:2021yqz,Bourget:2018obm, Favrod:2018xov} (see \cite{Gaume:2020bmp} for a comprehensive review).

In particular, in \cite{Badel:2019oxl} a semiclassical
method to determine the scaling dimensions of operators with large charge $Q$ was developed, offering UV complete realization of the effective field theory (EFT) approach considered in \cite{Hellerman:2015nra}. The authors computed the anomalous dimensions of the scalar operators $\phi^Q$ in the Abelian $\phi^4$-model at the Wilson-Fisher fixed point in $4-\epsilon$ dimensions. Since then, the method was generalized to several non-Abelian models with quartic interactions \cite{Antipin:2020abu, Antipin:2020rdw, Antipin:2021akb}, while Yukawa interactions were discussed in \cite{Sharon:2020mjs} in the context of Wess-Zumino model with cubic superpotential.

In this work, we generalize the method of \cite{Badel:2019oxl} to include fermionic interactions allowing us to present it (in section \ref{method}) in the streamlined ready-to-use form for generic Yukawa-quartic theory. As examples, we consider two models containing quartic,  and Yukawa interactions and compute contributions to the anomalous dimensions of the lowest lying charged operators. Our first example, in section \ref{Yukawa}, is Nambu-Jona-Lasinio-Yukawa (NJLY) model \cite{Nambu, Fei:2016sgs} containing Yukawa and quartic interactions and featuring Wilson-Fisher infrared fixed point in $D=4-\epsilon$ dimensions. For two fermion flavors $N_f=2$ and in $D=3$ this model describes the superconducting critical behaviour in graphene \cite{Roy:2012bor} while for $N_f = 1/2$ it is relevant for surface states of  topological insulators \cite{Zerf:2016fti, Lee:2006if, Grover:2013rc}. Intriguingly, the latter case has been conjectured to feature emergent supersymmetry at criticality.
Moreover, it is believed that the NJLY model provides a UV completion of the Nambu-Jona-Lasinio model in $2<D<4$ \cite{Zinn-Justin:1991ksq, Fei:2016sgs}.
Our second example, appearing in section \ref{Litim-Sannino}, is the model studied in \cite{Litim:2014uca} featuring a perturbative ultraviolet fixed point in $4$D. This theory provides a perturbative realization of the asymptotic safety scenario often considered in the search for a theory of quantum gravity \cite{Weinberg:1976xy, Niedermaier:2006wt} and serves as a basis for UV-complete BSM model building \cite{Bond:2017wut, Pelaggi:2017abg, Molinaro:2018kjz, Cacciapaglia:2018avr, Abel:2018fls, Hiller:2019mou}. Moreover, it offers the intriguing opportunity to study the large charge expansion in a non-supersymmetric four-dimensional CFT and, in this context, has been previously investigated in \cite{Orlando:2019hte}. Since in this model the scalar fields are not charged under the gauge group, the gauge fields will be spectators to the large charge dynamics. Interestingly, by varying the parameters of the theory, our calculation captures various regimes of the large-charge dynamics. In particular, we identify a perturbative limit where our results match diagrammatic computations, a generalized Gaussian phase characterized by a linear scaling of the lowest scaling dimension with the charge of the corresponding operators, and the superfluid phase captured by the large charge EFT.  
Section \ref{conclusions} summarizes our work and discusses future directions. Three appendices follow with complementary material about  details of the calculations in the main text.

\section{General Approach} \label{method}

Here we review and generalize the method of \cite{Badel:2019oxl} in order to compute the contributions to anomalous dimensions of the charged scalar operators from quartic $\lambda$ and  Yukawa $y$ interactions with couplings denoted collectively as  $\kappa_I\equiv\{\lambda_I, y_I^2 \}$ with index $I$ running over all the couplings. 
\begin{itemize}

\item {\it The method uses the power of conformal invariance which requires in the first step to tune our QFT to the perturbative fixed point of the renormalization group (RG). }

We require weakly-coupled fixed points (FPs) in  order to facilitate comparison to standard perturbative expansion with small coupling and, depending on the model, this perturbative FP can be of the Wilson-Fisher  or Banks-Zaks type. For  Wilson-Fisher type one engineers a small parameter $\epsilon$ by moving away from integer space-time dimension $d\to d-\epsilon$ while for the Banks-Zaks type the small parameter $\epsilon$ is built from the parameters of the model, such as number of field components, number of colors, flavors, etc. In general, the fixed point can be complex. At the end of the computation, we can invert the fixed point condition $\beta_I(\kappa_I^*, \epsilon)=0$ which gives the fixed point values of the couplings $\kappa_I^*$ as a function of $\epsilon$. Then we express our results for anomalous dimensions back as the power series in the couplings $\kappa_I^*$ and, taking the $\epsilon\to 0$ limit by simply removing the star symbol ($*$) from the couplings we obtain the results valid away from the FP. 

\item {\it Having CFT in flat space we map it to the cylinder $\mathbb{R}\times S^{d-1}$ using the fact that Weyl invariance at the fixed-point guarantees equivalence between the two theories. }

Technically, parametrizing $R^D$ and $\mathbb{R}\times S^{D-1}$ by $(r, \Omega^{D-1})$ and $(\tau, \Omega^{D-1})$ respectively the mapping is given by $r=R e^{\tau/R}$ where $R$ is the radius of the sphere. 
In addition, for the scalar fields the Weyl invariance requires adding the mass term $m^2 \bar{\phi} \phi$ where $m^2=(\frac{D-2}{2R})^2$ arising from the coupling to the Ricci scalar (see also Appendix \ref{sphere}).
For the fermions, starting with  the flat free   fermion theory $
S_f=\int d^Dx\,\bar\psi i\slashed\del\psi 
$
in $D$-dimensions, the curved space action is given by
\be
\label{curvedferm}
S^{cur}_f=\int_{\mathcal M} d^Dx\,\sqrt{-g}\,\,\bar\psi i\slashed\nabla_{\mathcal M}\psi \,,
\ee
where $\slashed\nabla_{\mathcal M}\equiv \g^{\m}\nabla_{\m}$ is the Dirac operator on  $\mathcal M$ manifold\footnote{In case of $S^D$ the eigenfunctions of the Dirac operator will be the corresponding $D$-dimensional spherical harmonics.}  and 
\be
\nabla_{\m}\psi=\del_{\m}\psi+\frac{1}{4}\omega_{\m}^{ab}\g_{ab}\,\psi\,\,.
\ee
The $\g$ matrices on curved background are related to the flat ones by the vielbein, namely $\g^{\m}=e^{\m}_{\hat\m}\g^{\hat\m}$ where $\m,\n,\dots$ is for the curved indices and $\hat\m,\hat\n,\dots$ is for the flat indices. The theory above enjoys Weyl invariance without extra modifications, as can be checked using  the Weyl transformations
\be
e^a_{\m}(x)\to e^{-\s(x)}e^a_{\m}(x)\,\,,\qquad \psi(x)\to e^{\frac{(D-1)}{2}\s(x)}\psi(x)\,.
\ee

\item {\it  On the cylinder we exploit the operator/state correspondence for the 2-point function of a scalar primary operator $\mathcal{O}$, using the fact that the flat space limit $x_i\to 0$ is equivalent to $\tau_i\to -\infty$ which relates the scaling dimension $\Delta_{\mathcal{O}}$ of the operator with the energy on the cylinder $E_\mathcal{O}$ as $E_\mathcal{O}=\Delta_{\mathcal{O}}/R$ i.e.
\be\langle \mathcal{O}^\dagger(x_f) \mathcal{O}(x_i)\rangle_{cyl}=|x_f|^{\Delta_\mathcal{O}}|x_i|^{\Delta_\mathcal{O}} \langle \mathcal{O}^\dagger(x_f) \mathcal{O}(x_i)\rangle_{flat}\stackrel{\tau_i\to -\infty}= e^{-E_\mathcal{O}(\tau_f-\tau_i)} \ . \ 
\ee}
To compute this energy $E_\mathcal{O}$ we choose an arbitrary state $|Q\rangle$ with a fixed charge $Q$ and evaluate the expectation value of the evolution operator $e^{-H T}$ ($T \equiv \tau_f-\tau_i$) in this state 
\be\langle Q |e^{-H T}|Q\rangle\stackrel{T\to \infty}\sim e^{-E_\mathcal{O}T} .
\ee
The $T\to\infty$ limit is saturated by the lowest energy and therefore $E_\mathcal{O}$ corresponds to the lowest energy eigenstate with charge $Q$.

 In practice we will fix the charge by imposing a charge-fixing condition and will compute the constrained path integral using semiclassical expansion around the vacuum of the fixed-charge theory which is determined by the chosen charge configuration. In this work, the vacuum will correspond to a superfluid phase with homogeneous charge density while for generalizations we refer to \cite{Hellerman:2017efx, Banerjee:2019jpw}. The charge configuration is defined by a set of $\mathcal{O}(1)$ parameters $\{q_i\}$ such that when we have multiple charges $Q_i$ corresponding to the Cartan generators of a non-Abelian group we take all of them to be large and of the same order, i.e. we rescale them as $Q_i = Q q_i$ with $1/Q$ being our small expansion parameter. The charges must correspond to the Cartan sub-algebra in order to be simultaneously observable. To introduce each of these charges into the grand canonical partition function we modify the Hamiltonian as $H\to H-\sum_i \mu_i Q_i$.

An interesting aspect of this approach is that by varying the charge configuration one can access the scaling dimension of a variety of different composite operators by performing a single computation. 
Another important feature of the computation is that a priori it only fixes the energy eigenvalues corresponding to a set of Cartan charges called weights, while the correspondence between the weights and the irreducible representations may require additional analysis 
\cite{Antipin:2021akb, Badel:2022fya}. 

\end{itemize}

We will work in the double-scaling limit
$
Q\to\infty, \ \kappa_I \to 0$ with  $Q \kappa_I= (\text{fixed})
$ as appropriate for the semiclassical expansion.
In the superfluid phase scalar bosons condense so that in the perturbative regime $Q \kappa_I\ll$ 1, the lowest energy eigenstate will correspond to the charged operator with minimal classical scaling dimension that can be built from the scalar fields. In the strongly coupled regime $Q \kappa_I\gg$ 1 the level crossing may occur hindering the identification of the lowest-lying operator. In any case, the semiclassical expansion of the lowest scaling dimension corresponding to the chosen charge configuration can be represented as
\be
E_Q R=\Delta_Q = \sum_{j=-1} \frac{1}{Q^j}\Delta_j \left(Q\kappa_I^*, \{q_i\}\right) \ .
\label{expansion}
\ee
Within each model we will compute the $\D_{-1}$ and $\D_{0}$ terms. Calculation of $\D_{-1}$ amounts to plugging the solutions of the classical equations of motion into the effective action while expanding around the classical solutions to quadratic order and performing the Gaussian integration we obtain the 1-loop contribution to the anomalous dimension $\D_0$. Since $\D_{-1}$ is the classical result it will be finite while calculation of $\D_{0}$ will require renormalization. Each $\Delta_{j}$ in eq.\eqn{expansion} resums an infinite number of Feynman diagrams with $\D_{-1}$ resumming the leading powers of the $Q$ at every loop order, $\D_{0}$ resumming the next-to-leading powers and so on. In fact, the usual perturbative loop expansion can be written as 
\be \label{PT}
\Delta_Q =Q\left(\frac{d-2}{2}\right) + \sum_{l=1} P_Q^{(l-\text{loop})} \qquad \text{where} \qquad P_Q^{(l-\text{loop})} = \sum_{k=0}^{l} C_{kl} Q^{l+1-k} \,,
\ee
where the coefficients $C_{kl}$ stem from the small-charge expansion of $\Delta_{k-1}$. In short, eq.\eqref{expansion} can be seen as a re-arraignment of conventional perturbation theory in eq.\eqref{PT}. At the $l$-th loop order one needs to determine $l+1$ coefficients $C_{kl}$. Since in practice we will only compute $\D_{-1}$ and $\D_{0}$, at the $l$-loops order we remain with $l-1$ unknown coefficients. However, these can be fixed by matching to the known perturbative results for  $l-1$ values of $Q$, if available. This procedure allows to "boost" perturbation theory with each extra loop-order of boosting requiring one additional input coefficient. 

Vice versa, expanding in the opposite large $Q\kappa_I$ limit one obtains the general form
\begin{align}
\Delta_{Q}= Q^{\frac{d}{d-1}}\left[\alpha_{1}+ \alpha_{2} Q^{\frac{-2}{d-1}}+\alpha_3 Q^{\frac{-4}{d-1}}+\ldots\right] +Q^0\left[\beta_0+ \beta_{1} Q^{\frac{-2}{d-1}}+\ldots\right] + \mathcal{O}\left( Q^{-\frac{d}{d-1}}\right)  \ ,
\label{largecharge}
\end{align}
which can be independently proven by building an EFT describing the superfluid phase realized by the large-charge sector of generic interacting CFT.
In this case, all the microscopic physics is encapsulated in parameters entering eq.\eqref{largecharge} \footnote{To be more precise, the UV physics is contained in the Wilson coefficients of the large-charge EFT from which eq.\eqref{largecharge} is derived.} with the exception, in odd dimensions, of $\beta_0$, which is universal and whose value is a robust prediction of the EFT formalism. In even dimensions one needs to include also $Q^p \log(Q)$ terms, with $p \le 0$ to be determined, which stem from the cancellation of the UV divergences and have universal coefficients \cite{Cuomo:2020rgt}. The most relevant examples of theories that do not satisfy eq.\eqref{largecharge} are given by free scalar theories and BPS operators in supersymmetric theories with scaling dimension  $\Delta_Q = Q \Delta_{Q=1}$.

\section{NJLY model}

\label{Yukawa}

 In this section we consider the multi-flavor NJLY model \cite{Nambu, Fei:2016sgs} containing Yukawa and quartic interactions, at the Wilson-Fisher infrared FP in $4-\epsilon$ dimensions. The interesting features of NJLY-model are that it is the UV completion of NJL model and in the limit of a single Majorana fermion, it leads to the supersymmetric Wess-Zumino model of a single chiral superfield. 

Our starting point is the NJLY Lagrangian
\be \label{NJLYLAG}
\mathcal L _\text{NJLY}=\frac{1}{2}(\del_{\m}\phi_1)^2+\frac{1}{2}(\del_{\m}\phi_2)^2+\bar\psi_j\slashed\del\psi^j+g\,\bar\psi_{Rj}\bar\phi\psi_L^j+g\,\bar\psi_{Lj}\phi\psi^j_R+\frac{(4 \pi)^2\l}{24}\left(\bar\phi\phi\right)^2 \,,
\ee
where $j=1,\dots,N_f$ is the flavor index of the Dirac fermions and $\phi = \phi_1+ i \phi_2$ is a complex scalar field. The model enjoys a $U(1)$ chiral symmetry expressed in the form
\be
\label{axial}
\phi\to e^{-2i\a}\phi,\qquad  \psi_{Lj}\to e^{-i\a}\psi_{Lj},\quad \psi_{Rj}\to e^{i\a}\psi_{Rj}\,.
\ee
We now follow the steps outlined in Sec.\ref{method} starting with the fact that in $D=4-\e$ dimensions the NJLY model features an infrared fixed point of the Wilson-Fisher type at 
\begin{align} \label{NJLYFP}
   \frac{ g^{2*}}{(4 \pi)^2} &=  \frac{\e}{4 (1+ N_f)} + \frac{ \left(-4 N_f^2+448 N_f+4 \sqrt{N_f \left(N_f+38\right)+1} \left(N_f+19\right)-274\right)}{1600 \left(N_f+1\right){}^3} \epsilon ^2 + \mathcal{O}\left(\e^3 \right) \nonumber\,, \\ 
  \l^* &= \frac{3 \left(\sqrt{N_f \left(N_f+38\right)+1}-N_f+1\right)}{20 \left(N_f+1\right)} \e + \frac{9 \epsilon ^2 }{2000 \left(N_f+1\right){}^3 \sqrt{N_f \left(N_f+38\right)+1}} \nonumber \\ & \times \bigg(\sqrt{N_f \left(N_f+38\right)+1} \left(3 N_f \left(-4 N_f^2+38 N_f+161\right)+20\right)+20 + 12 N_f^4+114 N_f^3  \nonumber \\ &  +191 N_f^2-1637 N_f\bigg) + \mathcal{O}\left(\e^3 \right) \ .
\end{align}

Mapping this CFT to the cylinder $\mathcal M=\mathbb{R}\times S^{D-1}$, the action reads
\begin{align}
S _\text{NJLY}=\int d^Dx\sqrt{-g}\bigg(&\frac{1}{2}(\del_{\m}\phi_1)^2+\frac{1}{2}(\del_{\m}\phi_2)^2+\frac{1}{2}m^2\phi_1^2 +\frac{1}{2}m^2\phi_2^2+\bar\psi_j\slashed\nabla_{\mathcal M}\psi^j \nonumber \\ &  +g\,\bar\psi_{Rj}\bar\phi\psi_L^j+g\,\bar\psi_{Lj}\phi\psi^j_R+\frac{(4 \pi)^2\l}{24}\left(\bar\phi\phi\right)^2\bigg) \,.
\end{align}
 By using the operator/state correspondence for a charged state denoted by $|Q\rangle$, we compute the expectation value of the evolution operator in the infinite time limit as
\be
\label{amplitude}
 \langle Q| e^{-HT }|Q\rangle=\mathcal Z^{-1}\int \mathcal D\chi_i\mathcal D\chi_f e^{-i\frac{Q}{R^{d-1}\Omega_{d-1}}\left[\int d\Omega_{d-1}(x_f-x_i)\right]}\int_{(\rho,\chi)=(f,\chi_i)}^{(\rho,\chi)=(f,\chi_f)} \mathcal D\rho\mathcal D\chi \mathcal D \bar\psi \mathcal D \psi \,e^{-S} \,,
\ee
where we defined
\be
\mathcal Z=\int \mathcal D \phi\mathcal D\bar\phi \mathcal D \bar\psi \mathcal D \psi\,e^{-S}\,\,,
\ee
and 
\be
\label{phirhochi}
\phi(x)=\rho(x)\,e^{i\chi(x)}\,\,,\qquad \bar\phi(x)=\rho(x)\,e^{-i\chi(x)} \ .
\ee 
Note that the term 
\be
\int d\Omega_{d-1}(\chi_f-\chi_i)=\int_{-T/2}^{T/2}d\tau\int d\Omega_{d-1}\dot\chi
\ee
is the charge fixing condition. Then the two-point function can be written in the form
\be\label{2point}
 \langle Q| e^{-HT }| Q\rangle=\mathcal Z^{-1}\int_{\rho=f}^{\rho=f} \mathcal D\rho\mathcal D\chi \,e^{-S_{eff}} \,,
\ee
where
\be
\label{effrhochi}
\begin{split}
S_{eff}=\int_{-T/2}^{T/2}d\tau\int d\Omega_{D-1}&\Big[\frac{1}{2}(\del\rho)^2+\frac{1}{2}\rho^2(\del\chi)^2+\frac{m^2}{2}\rho^2+\frac{(4 \pi)^2\l}{24}\rho^4+\frac{i\,Q}{R^{d-1}\Omega_{d-1}}\dot\chi\\
&+\bar\psi_j\slashed\nabla_{\mathcal M}\psi^j+g\,\bar\psi_{Rj}\psi_L^j\, \rho e^{-i\chi}+g\,\bar\psi_{Lj} \psi^j_R\,\rho \,e^{i\chi}\Big] \ .
\end{split}
\ee
To compute the leading $\D_{-1}$ term in the semiclassical expansion we need to solve the classical equations of motion supplemented by equation fixing the value of the charge
\begin{align}
\label{eomrho}
&-\nabla^2\rho+\left[(\del\chi)^2+m^2\right]\rho+\frac{(4 \pi)^2\l}{6}\rho^3+g\,\bar\psi_{Rj}\psi_L^j\,  e^{-i\chi}+g\,\bar\psi_{Lj} \psi^j_R\, \,e^{i\chi}=0\,,\\
\label{eomchi}&-\nabla_{\m}\left(\rho^2g^{\m\n}\del_{\n}\chi\right)+g\,\bar\psi_{Rj}\psi_L^j\,\rho  e^{-i\chi}+g\,\bar\psi_{Lj} \psi^j_R\, \rho\,e^{i\chi}=0\,,\\
\label{eomcharge}&i\rho^2\dot\chi=\frac{Q}{R^{d-1}\Omega_{d-1}} \,,
\end{align}
where $-\nabla^2=-\del_{\tau}^2-\nabla_{S^{D-1}}^2$ (for details see Appendix \ref{sphere}). The fermionic  equations of motion are not written since they have trivial solution $\psi^{cl}_{L,R}=0$. Plugging these into  eq.\eqn{eomrho}-\eqn{eomcharge} the extra fermionic terms will vanish leaving us with the classical bosonic equations of motion for the complex scalar field. At the classical level, this is exactly the computation considered in \cite{Badel:2019oxl} so we will be brief. Choosing the ground state corresponding to the superfluid phase with homogeneous charge density as
\be
\label{sol}
\rho=f,\qquad\chi=-i\m\tau \,,
\ee
and substituting this ansatz 
(together with $\psi^{cl}_{L,R}=0$) in \eqn{effrhochi} we obtain the classical effective action
\be
\label{Seff}
\frac{S_{eff}}{T}=\frac{Q}{4}\left(3\m+\frac{m^2}{\m}\right)\,.
\ee
Note that from eq.\eqn{sol} and the equations of motion above we have
\be
\label{chem}
\m^3-\m=\frac{4}{3}\l Q\,\,.
\ee
Setting $D=4$ in eq.\eqn{chem} the chemical potential can be expressed in terms of the 't Hooft coupling as
\begin{equation}
  \mu = \frac{3^\frac{1}{3}+\left(x + \sqrt{-3+x^2}\right)^\frac{2}{3}}{3^\frac{2}{3}\left(x + \sqrt{-3+x^2}\right)^\frac{1}{3}} \,, \qquad \qquad x \equiv 6 \l  Q  \,.
  \label{fourmu}
\end{equation}
Substituting this into \eqn{Seff}, the leading order in the semiclassical expansion reads
\be
 \label{classic}
 4\,\Delta_{-1}=  \frac{3^\frac{2}{3}\left(x+\sqrt{-3+x^2}\right)^{\frac{1}{3}}}{3^\frac{1}{3}+\left(x+\sqrt{-3+x^2}\right)^{\frac{2}{3}}}  + \frac{3^\frac{1}{3}\left(3^\frac{1}{3}+\left(x+\sqrt{-3+x^2}\right)^{\frac{2}{3}}\right)}{\left(x+\sqrt{-3+x^2}\right)^{\frac{1}{3}}}  \ .
\ee
Having obtained the classical contribution to the anomalous dimension, we now proceed to the evaluation of the first quantum correction, namely $\D_0$. 
Therefore, we expand the scalar fields around the non-trivial saddle point solution above 
\be
\label{rhochi}
\rho(x)=f+r(x),\qquad \chi(x)=-i\m\tau+\frac{1}{f}\,\pi(x) \,,
\ee
as well as expand the fermion fields around the zero classical solutions. In addition, in order to get rid of the unwanted phases in the Yukawa interaction terms, we redefine the fermions as
\be
\label{psimu}
\psi_{L}\to {\psi_L} \,e^{\m\tau/2}\,\,,\qquad\psi_R\to {\psi_R}\,e^{-\m\tau/2} \,,
\ee
and note that under this change the fermion kinetic term shifts as
\be
\bar\psi_j\slashed\nabla_{\mathcal M}\psi^j\to \frac{\m}{2}\, \bar\psi_{j}\g^0\psi^j+\bar\psi_{j}\slashed\nabla_{\mathcal M}\psi^j\,\,.
\ee
Plugging the expressions \eqn{rhochi} and \eqn{psimu} into the action \eqn{effrhochi} and keeping only quadratic terms in the fluctuations, we arrive at
\be
\label{s2}
\begin{split}
S^{(2)}=\int_{-T/2}^{T/2}d\tau\int d\Omega_{d-1}&\Big[\frac{1}{2}(\del r)^2+\frac{1}{2}(\del\p)^2-2i\m\, r\del_{\tau}\p+(\m^2-m^2)r^2\\
&+\frac{\m}{2}\, \bar\psi_{j}\g^0\psi^j+\bar\psi^j\slashed\nabla_{\mathcal M}\psi^j+g\,f\bar\psi_{Lj}\psi_R^j+g\,f\bar\psi_{Rj} \psi^j_L\Big] \ .
\end{split}
\ee
The Gaussian  integral  of the action \eqn{s2} on $\mathbb{R}\times S^3$ is cast in the form
\be
\int \mathcal Dr\mathcal D\pi \mathcal D\bar\psi\mathcal D\psi\,e^{-S^{(2)}}=\frac{\det\, F}{\det\, B} \,,
\ee
where $F$ denotes the fermionic determinant and $B$ the bosonic one. For our evaluation we need the eigenvalues of the Laplacian on $S^3$. These are given in Appendix \ref{sphere} including their degeneracies on $S^D$. Here we just recall that $(J_{\ell(s)}^2,n_b)$ denote  the boson eigenvalues and degeneracies respectively, while for fermion these are denoted by $(\l_{f\pm},n_f)$. 
The dispersion relations of the scalar modes are
\begin{equation} \label{confradio}
    \omega_{\pm}(\ell) = \sqrt{J^2_{\ell(s)}+3\mu^2-m^2\pm \sqrt{4 J^2_{\ell(s)}\mu^2+\left(3\mu^2-m^2\right)^2}}  \,,
\end{equation}
and describe a massive mode with mass $\omega_+(0)=6 \mu^2-2 m^2$ and a massless mode with speed $c_S=\sqrt{\frac{\mu^2-m^2}{3\mu^2-m^2}}$. The latter is the Goldstone boson stemming from the spontaneously broken $U(1)$ symmetry. The dispersion relations of the fermion modes are
\begin{equation} \label{confradioferm}
    \omega_{f\pm}(\ell) = \sqrt{\frac{3 g^2 \left(\mu ^2-m^2\right)}{8 \pi ^2 \l}+\left(\frac{\mu }{2}+\l_{f\pm}\right)^2}\,\,.
\end{equation}
The presence of Yukawa interactions destroys the Fermi surface existing in the free case at $\l_{f\pm}=-\mu/2$ implying that in order  to compute the fermionic contribution to $\Delta_0$, we only sum the zero-point energies so that altogether we have \footnote{Recall that in flat-space the Fermi surface is defined by the solution $p_0=\omega(p)=0$ for some $p=p_{F}>0$. In presence of Fermi surface and following the textbook flat-space calculation of the free energy $\Omega=H-\mu Q$ we expect to have an additional "matter" contribution $\Omega_{matter}=-2 \int \frac{d^3 p}{(2 \pi)^3} (\mu-\omega(p))\theta(\mu-\omega(p))$ where $w(p)$ is the corresponding flat space dispersion relation. To elucidate the role of this term, it would be interesting  to consider fermionic systems with Fermi sea \cite{Cherman:2013rla}. }
\begin{equation}
\label{eq:one-loop-det1}
\Delta_0 = \frac{1}{2}\sum_{\ell=0}^\infty \left[n_{b}(\ell)(\omega_+(\ell)+\omega_-(\ell))-N_f n_{f}(\ell)(\omega_{f+}(\ell)+\omega_{f-}(\ell))\right]\,\,.
\end{equation}
The sum over $\ell$ needs to be regularized. We perform regularization by subtracting the divergent powers of $\ell$ in the expansion of the summand around $\ell=\infty$. The sum over the subtracted terms is then regularized with the corresponding zeta function value. The procedure allows isolating a $1/\epsilon$ pole in dimensional regularization stemming from $\sum_\ell \ell^{D-5} = \zeta(1+\epsilon) = \frac{1}{\e} + \gamma_E + \mathcal{O}(\e)$ with $\gamma_E$ the Euler-Mascheroni constant. This pole is then canceled by a corresponding contribution in the renormalization of $\Delta_{-1}$ which is performed by expanding the bare couplings in powers of the renormalized ones and then tuning the latter to their FP values to remove the dependence on the arbitrary renormalization scale. Since the two $1/\e$ terms arising in the procedure stem from two different orders of the semiclassical expansion, namely $\Delta_{-1}$ and $\Delta_0$, their cancellation provides a useful self-consistency check of our calculation.
The renormalized result reads
\begin{align} \label{square}
\Delta_0 =  \Delta_0^{(b)} - N_f \Delta_0^{(f)} \,,
\end{align}
where
\begin{align} \label{square1}
\Delta_0^{(b)}(\l^* \bar Q)=  -\frac{15 \mu^4  +6 \mu^2  -5}{16}
+\frac{1}{2} \sum_{\ell=1}^\infty\sigma^{(b)}(\ell)
+\frac{\sqrt{3\mu^2-1}}{\sqrt{2}} \,,
\end{align}
and 
\begin{align} \label{1looferm}
  \Delta_0^{(f)}(\l^* \bar Q, g^*) &= -6 -\frac{3 g^2 \left(\mu ^2-1\right) \left(g^2 \left(9 \mu ^2+3\right)+8 \pi ^2 \l \left(13-3 \mu ^2\right)\right)}{512 \pi ^4 \l^2} +\frac{1}{2} \sum_{\ell=1}^\infty\sigma^{(f)}(\ell) \nonumber \\ &+\frac{1}{\sqrt{2} \pi }\left(\sqrt{\frac{3 g^2 \left(\mu ^2-1\right)}{\l}+2 \pi ^2 (\mu -3)^2}+\sqrt{\frac{3 g^2 \left(\mu ^2-1\right)}{\l}+2 \pi ^2 (\mu +3)^2} \right) \ .
\end{align}
The summands appearing in the expressions above are given by
\begin{align}
\sigma^{(b)}(\ell) =(1+ \ell)^2 \left[ \omega_+ (\ell) + \omega_-(\ell) \right]  -2 \ell^3-6  \ell^2-2 \mu ^2-2 \left(\mu ^2+2\right)  \ell+\frac{5 \left(\mu ^2-1\right)^2}{4  \ell} \,,
\end{align}
and
\begin{align}
\sigma^{(f)}(\ell) &=2 (1+ \ell)(2+ \ell)[\omega_{f+}(\ell)+\omega_{f-}(\ell)]-2 (\ell+1) (\ell+2) (2 \ell+3) \nonumber \\ & -\frac{3 g^2 \left(\mu ^2-1\right) \left(4 \ell^2+6 \ell+\mu ^2-1\right)}{16 \pi ^2 \l \ \ell} + \frac{9 g^2 \left(\mu ^2-1\right)^2}{128 \pi ^4 \l^2 \ \ell} \,\,.
\label{1loofermsum}
\end{align}
Eqs.\eqref{1looferm} and \eqref{1loofermsum} constitute the main result of this section. As mentioned, we can make contact with diagrammatic computations by expanding our results in the small $Q \e$ limit. The expansion for $\Delta_{-1}$ and $\Delta_0^{(b)}$ can be found in \cite{Badel:2019oxl} and we do not repeat it here. Instead, we report below the small $Q \e$ expansion of $\Delta_0^{(f)}$ up to the three-loops order
\begin{align}
 \Delta_0^{(f)} &= Q \left(\frac{g^2}{8 \pi ^2}-\frac{3 g^4}{32 \pi ^4 \lambda } \right) + Q^2 \left(\frac{g^2 \lambda }{12 \pi ^2}-\frac{g^4}{32 \pi ^4}\right) + Q^3 \left(\frac{g^6 \zeta (3)}{64 \pi ^6}-\frac{g^2 \lambda ^2}{18 \pi ^2} +g^4 \lambda \frac{  1-3 \zeta (3)}{48 \pi ^4} \right)  \,.
\end{align}
In App.\ref{riscomp}, we provide explicit results for the full $\Delta_Q$ up to the $6$-loops order, in a form suited for comparisons with diagrammatic calculations. By rewriting the small-charge expansion of $\Delta_0$ and $\Delta_{-1}$ at the fixed point \eqref{NJLYFP} we obtain $\Delta_Q$ to order $\mathcal O(\e^2)$
\begin{align} \label{booster}
\Delta_Q &= Q + \left[\frac{Q \left(-\sqrt{N_f \left(N_f+38\right)+1}+N_f-11\right)}{20 \left(N_f+1\right)} + \frac{Q^2 \left(\sqrt{N_f \left(N_f+38\right)+1}-N_f+1\right)}{20 \left(N_f+1\right)} \right] \e \nonumber \\ &+ \left[\left(\textcolor{red}{\frac{64 N_f^4+3748 N_f^3+10557 N_f^2+5581 N_f-50}{2000 \left(N_f+1\right){}^3 \sqrt{N_f \left(N_f+38\right)+1}}-\frac{32 N_f^3+391 N_f^2+1797 N_f+25}{1000 \left(N_f+1\right){}^3}}        \right) Q  \right. \nonumber \\ & +  \left. \left( \frac{-84 N_f^4-4318 N_f^3-3327 N_f^2-4251 N_f+80}{2000 \left(N_f+1\right){}^3 \sqrt{N_f \left(N_f+38\right)+1}}+ \frac{84 N_f^3+1722 N_f^2+2729 N_f+80}{2000 \left(N_f+1\right){}^3}\right) Q^2  \right. \nonumber \\ & -  \left. \left(\frac{\left(1-N_f\right) \sqrt{N_f \left(N_f+38\right)+1}+N_f^2+18 N_f+1}{100 \left(N_f+1\right){}^2}\right) Q^3 \right] \e^2 + \mathcal{O}\left( \e^3\right) \,.
\end{align} 
The fixed charge operator corresponding to this result is $\phi^Q$ and
$\Delta_{-1}$ and $\Delta_0$ resum, respectively, the leading and next to leading powers of $Q$ at every order in $\e$. The red term in eq.\eqref{booster} would stem from the small charge expansion of $\Delta_1$ and does not follow directly from our computation. We instead fixed it by matching the result to  
the known $2$-loop scaling dimension of $\phi$ (see \cite{Fei:2016sgs}), which corresponds to $Q=1$. We have checked that for $Q=2$ our result reproduces the known $2$-loop scaling dimension of the operator $\phi \phi$ (see \cite{Fei:2016sgs}).

The large 't Hooft coupling limit of $\Delta_{-1}$ can be trivially obtained from eq.\eqref{classic}. In order to expand $\Delta_0$, we follow \cite{Cuomo:2021cnb} and split the sum over $\ell$ in eq.\eqref{square1} as
\be
\sum_{\ell=1}^\infty \sigma^{(b)} = \sum_{\ell=1}^{\L \mu} \sigma^{(b)}+\sum_{\L \mu+1}^\infty \sigma^{(b)} \,,
\ee
where $\L$ is an arbitrary cutoff scale such that $\L \mu$ is an integer. The sum over the low modes ("low $\ell$") can be computed by expanding the summand for large $\mu$ and then computing the sum over $\ell$. We have
\be
\frac{1}{2} \sum_{\ell=1}^{\L \mu} \sigma^{(b)} = \frac{5}{8} \mu ^2  \left(\mu ^2-2\right) H_{\L \mu } - \frac{1}{2} \mu ^2 \left(\L \mu  (\L \mu +3)\right) + \cO \left(\mu^0 \right) \,,
\ee
where $H_{\L \mu}$ is the $(\L \mu)^{\text{th}}$ Harmonic number. The  polynomial terms in $\L$ can be neglected since we are ultimately interested in taking the limit $\L \to 0$. The Harmonic number can be expanded for large $\L \mu$ obtaining
\be \label{low}
\frac{1}{2} \sum_{\ell=1}^{\L \mu} \sigma^{(b)} = \frac{5}{8} \mu ^2 \left(\mu ^2-2\right) (\log \mu +\gamma_E ) +\frac{5}{8} \mu ^2 \left(\mu ^2-2\right) \log \L +\frac{5 \mu ^3}{16 \L}-\frac{5 \mu ^2}{96 \L^2}-\frac{5 \mu }{8 \L} + \cO \left(\mu^0 \right)\,.
\ee 
To evaluate the sum over the high modes we introduce $k \equiv \ell/\mu$ and make use of the Euler-Maclaurin summation formula
\be \label{eul}
\frac{1}{2} \sum_{\L \mu+1}^\infty \sigma^{(b)}  ={\frac{\mu}{ 2}}\int_{\L}^\infty dk \ \Sigma(k)-{\frac{\Sigma(\L)}{ 4}}-\sum_{m=1}{\frac{B_{2m}}{ 2 (2m)! (\mu)^{2m-1}}}\, \Sigma^{(2m-1)}(\L)\,, \quad
\Sigma(k)\equiv \sigma^{(b)}(k\mu)\,.
\ee
The above is evaluated by expanding $\Sigma(k)$ as $\Sigma(k) =\mu^3 \left( \Sigma_1(k) + \frac{1}{\mu} \Sigma_2(k) + \dots \right)$ and then expanding \eqref{eul} for small $\L$ keeping only the terms that do not vanish in the $\L \to 0$ limit. In the procedure, the integrals over the $\Sigma_i(k)$ need to be regularized by performing subtractions that regularize the infrared behaviour of the integrands leaving untouched their UV asymptotics. We obtain 
\begin{align} \label{high}
 \frac{1}{2} \sum_{\L \mu+1}^\infty \sigma^{(b)} &= \mu^4 \left(\frac{5}{16}  \log \left(\frac{8}{5}\right) + \int_0^\infty \Sigma_1^{\text{reg}}(k) d k -\frac{5}{8} \log \L \right) -\frac{5 \mu^3}{16 \L}  \nonumber \\ 
 \nonumber&+ \mu^2 \left(\frac{1}{24}  \left(14+15 \log \left(\frac{5}{4}\right)\right) + \int_0^\infty \Sigma_3^{\text{reg}}(k) d k + \frac{5}{4} \log \L + \frac{5 \mu ^2}{96 \L^2} \right)\\
 &+ \mu\left( \frac{5 }{8 \L} - \sqrt{\frac{3}{2}} \right) \,,
\end{align}
where the expression of the integrands is given in App.\ref{App1}.
Combining \eqref{high} and \eqref{low}, all the $\L$-dependent terms drop leaving us with
\begin{align}
\Delta_0^{(b)} &=\frac{5}{8} \mu ^2 \left(\mu ^2-2\right)\log \mu+\mu^4 \left( \frac{5}{16} \left(2 \gamma_E -3+\log \left(\frac{8}{5}\right)\right) + \int_0^\infty \Sigma_1^{\text{reg}}(k) d k \right) \nonumber \\ & + \mu^2 \left(\frac{5}{24} \left(-6 \gamma_E +1+3 \log \left(\frac{5}{4}\right)\right)+ \int_0^\infty \Sigma_3^{\text{reg}}(k) d k\right) + \cO\left(\mu^0 \right) \,.
\end{align}
We have checked that the above is consistent with the results of \cite{Badel:2019oxl}, which have been obtained via a numerical fit to $\Delta_0^{(b)}$. 

The same procedure can be applied to $\Delta_0^{(f)}$, obtaining
\begin{align}
\Delta_0^{(f)} &= \frac{\mu ^4}{3072 \pi ^4 \l^2}\bigg(27 g^4 (4 \gamma_E -5-28 \log (2))+18 g^2 \left(3 g^2-8 \pi ^2 \l\right) \log \left(\frac{1536 g^2 \mu ^2}{\pi ^2 \l}\right) \nonumber \\ &  +144 \pi ^2 g^2 \l (-2 \gamma_E +1+14 \log (2))+32 \pi ^4 \l^2\bigg)-\frac{\mu ^2}{256 \pi ^4 \l^2} \bigg(18 \gamma_E  g^4-48 \gamma_E  \pi ^2 g^2 \l  \nonumber \\ &-20 \pi ^2 g^2 \l+16 \pi ^4 \l^2 +3 g^2 \left(3 g^2-8 \pi ^2 \l\right) \log \left(\frac{3 g^2 \mu ^2}{32 \pi ^2 \l}\right)\bigg) + \cO\left( \log \mu \right) \,.
\end{align}
At the fixed point \eqref{NJLYFP}, the large 't Hooft coupling expansion of $\Delta_{-1}$, $\Delta_0$ matches the general non-perturbative form \eqref{largecharge}, which we rewrite as
\be
\Delta_Q = \frac{1}{\epsilon }(Q \epsilon )^{\frac{4-\epsilon }{3-\epsilon }}\left[\alpha_{10}+ \alpha_{11} \epsilon +\cO\left(\e^2\right) \right] +\frac{1}{\epsilon }(Q \epsilon )^{\frac{2-\epsilon }{3-\epsilon }}\left[\alpha_{20}+ \alpha_{21} \epsilon +\cO\left(\e^2\right)  \right] + \cO\left(Q^0\right) \,,
\ee
making evident that we are computing the $\alpha_i=\alpha_{i 0}+\alpha_{i 1}\epsilon+\dots$ coefficients of eq.\eqref{largecharge} in the $\e$-expansion. The leading/next-to-leading orders of the latter, stem respectively from $\Delta_{-1}$ and $\Delta_0$. We can now take $\e \to 1$ in order to obtain predictions for the $\alpha_i$ coefficients of the three-dimensional theory. The latter may be in future compared either with experiments or Monte-Carlo simulations along the lines of \cite{Banerjee:2017fcx, Banerjee:2019jpw, Banerjee:2021bbw, Singh:2022akp}. Our results for $N_f=1/2, 1, 2, 3, 4, 5$ are summarized in \autoref{heplat}. Notice that the NLO correction is small for $\alpha_1$ but quite large for $\alpha_2$ as already noted in the $g=0$ case \cite{Badel:2019oxl}.

\begin{table}[t]
\begin{center}
\begin{tabular}{|c|c|c|c|c|}
\hline
&\multicolumn{2}{c|}{$\alpha_1$} & \multicolumn{2}{c|} { $\alpha_2$} \\
\hline
 & LO & NLO & LO &  NLO\\ \hline
 $N_f = 1/2$ & 0.655  & 0.545 & 0.572   & 0.217 \\
\hline
$N_f = 1$ & 0.644 & 0.596 & 0.582  & 0.244  \\
\hline
$N_f = 2$ & 0.608  & 0.595  &  0.617 &  0.312  \\
\hline
$N_f = 3$ & 0.578  & 0.567  &  0.649  & 0.377  \\
\hline
$N_f = 4$ & 0.553 &  0.536 & 0.679  & 0.435 \\
\hline
$N_f = 5$ & 0.532   & 0.507 & 0.705  &  0.488  \\
\hline
\end{tabular}
\end{center}
\caption{\label{heplat}
Value of the coefficients $\alpha_1$ and $\alpha_2$ in \eqref{largecharge} computed to LO and NLO in the $\e$-expansion.}
\end{table}


\subsection{Emergent supersymmetry - The Wess-Zumino model}

In four dimensions and for $N_f=1/2$ and $(4 \pi)^2 \l =3 g^2 \equiv \frac{3}{2} \kappa^2$, the Lagrangian of the NJLY model \eqref{NJLYLAG} reduces to the Wess-Zumino model describing a four-component Majorana fermion and a complex scalar \cite{Wess:1973kz, Lee:2006if}
\be
\mathcal{L}_{WZ}= \frac{1}{2}(\del_{\m}\phi_1)^2+ \frac{1}{2}(\del_{\m}\phi_2)^2+\frac{1}{2}\bar\psi\slashed\del\psi+\frac{\kappa}{2 \sqrt{2}}\,\bar\psi(\phi_1+i \gamma_5 \phi_2)\psi+\frac{\kappa}{16}\left(\phi_1^2+\phi_2^2\right)^2\,\,.
\ee
This is an $\mathcal{N}=4$ supersymmetric theory of a single chiral superfield $\Phi$ and superpotential $\mathcal{W}=\kappa \ \Phi^3$. The chiral symmetry of the NJLY model becomes the $R$-symmetry of the Wess-Zumino theory with the $R$-charge related to the axial one as $R_{\phi}=\frac{2}{3}Q$.  In $D=4-\e$ the fixed point in the $\k$ coupling occurs at
\be
\frac{\k^{*2}}{(4 \pi)^2}=\frac{\epsilon }{3}+ \frac{\epsilon ^2}{9} +\frac{1}{36} (1-4 \zeta (3)) \epsilon ^3 + \mathcal{O}(\e^4)\,.
\ee
At this fixed point, the chiral operator $\phi$ has conformal dimension protected by supersymmetry and determined by its $R$-charge as
\be \label{dimphi}
 \Delta_\phi = \frac{D-1}{2}R_\phi = \frac{D-1}{3}\,\,.
\ee
Moreover, due to the equations of motion, $\phi^2$ is a descendant of $\phi$ and thus $\Delta_{\phi^2}=\Delta_{\phi}+1$.
This relation and eq.\eqref{dimphi} are in agreement with eq.\eqref{booster} for $N_f=1/2$. In the same way, all the results in the previous section apply here when $N_f=1/2$ and the couplings are tuned to the supersymmetric fixed-point. Moreover, since the scaling dimensions for $Q=1$ and $Q=2$ are now known to all orders of the $\e$-expansion, we can use our semiclassical results to derive $\Delta_Q$ at order $\e^3$. We find
\be
\Delta_Q =Q +\frac{1}{6} (Q-3) Q \epsilon-\frac{1}{18} (Q-2) (Q-1) Q \epsilon ^2 +\frac{1}{108} (Q-2) (Q-1) Q (4 Q -11+ 12 \zeta (3)) \epsilon ^3\,\,.
\ee
Finally, further evidence for emergent supersymmetry in the critical NJLY model can be obtained by looking at the dispersion relations of the fermions \eqref{confradioferm}, which for $N_f=1/2$ reduce to
\begin{equation}
    \omega_{f\pm}(\ell) = \sqrt{2(\mu^2-m^2)+\left(\frac{\mu }{2}+\l_{f\pm}\right)^2} \approx  \frac{3 \mu }{2}+\frac{\l_{f\pm}}{3} + \mathcal{O}(\l_{f\pm}^2)\,.
\end{equation}
As discussed in \cite{Hellerman:2015nra}, the value of the mass of the fermions is dictated by Bose-Fermi degeneracy to be $\frac{3}{2}\dot{\chi} =\frac{3}{2}\mu$. Analogously, the speed of the fermion modes takes the value $\pm 1/3$ as a consequence of supersymmetry.

\section{An asymptotically safe model in $D=4$ }
\label{Litim-Sannino}

We now proceed to the next model generalizing the NJLY model above in several aspects. First, we wish to study CFT in exactly four dimensions which can be achieved by introducing gauge fields. Specifically, the model enjoys $SU(N_c)$ gauge invariance and contains $N_f$ flavors of Dirac fermions $\Psi_i$ in the fundamental of $SU(N_c)$ plus an $N_f \times N_f$ complex matrix scalar field $\Phi$. This  field  transforms in the $(N_f, \bar{N_f})$ representation of the $U(N_f)\times U(N_f)$ symmetry and can be written in terms of $2N_f^2$ real scalar fields:
\be
(\Phi)_{a\alpha}=
\frac{\phi+i\eta}{\sqrt{2N_f}}\delta_{a\alpha}+\sum_{A=1}^{N_f^2-1} (h^A+i\pi^A)T^A_{a \alpha} \ .
\label{matrixH}
\ee
The scalar field $\Phi$ is not charged under the gauge group and therefore, as we will see, the gauge fields up to  NLO in the large charge expansion in this model will be spectators. In addition, the group-theoretical structure of this model has a more rich structure and this allows us to illustrate the procedure of identifying the irreducible representation from the given charge configuration. Finally, the model has two scalar couplings which will lead to some technical differences from NJLY model.

The Lagrangian reads \cite{Litim:2014uca} 
\begin{align}\label{eq:fullLag}
    \mathcal{L} =& -\frac{1}{2}\Tr (F^{\mu\nu}F_{\mu\nu}) + \Tr(\bar \Psi i\slashed{D} \Psi) + y \,\Tr(\bar \Psi_L \Phi \Psi_R + \bar \Psi_R \Phi^\dagger \Psi_L) \nonumber \\ &+ \Tr(\del_\mu \Phi^\dagger \del^\mu \Phi ) - u\left[\Tr(\Phi^\dagger \Phi)\right]^2 - v\,\Tr(\Phi \Phi^\dagger \Phi \Phi^\dagger )\,\,.
\end{align}
In the Veneziano limit 
\be
N_f \to \infty \,, \qquad N_c \to \infty \,, \qquad z \equiv \frac{N_f}{N_c}= \text{ fixed} \,, \label{Veneziano}
\ee
this theory displays a perturbative UV fixed-point for small $\delta \equiv z-\frac{11}{2}$, which at the leading order reads \cite{Litim:2014uca} \footnote{There is another real fixed point for the coupling $\alpha_v$ given by $\alpha_{v2}^* =-\frac{1}{19} \left(\sqrt{20+6 \sqrt{23}}+2 \sqrt{23}\right) \delta$ which however leads to an unbounded from below scalar field potential \cite{Litim:2014uca}.} 
\be \label{LSFP}
\begin{split} 
&\alpha_g^* = \frac{26}{57}\delta \,, \quad \alpha_y^* = \frac{4}{19}\delta  \,, \quad \alpha_h^* = \frac{\sqrt{23}-1}{19}\delta  \,,\\
&\alpha_{v1}^* =\frac{1}{19} \left(\sqrt{20+6 \sqrt{23}}-2 \sqrt{23}\right) \delta\,,
\end{split}
\ee
where we introduced the rescaled couplings appropriate for the Veneziano limit
\begin{align}\label{eq:couplings}
  \alpha_g & = \frac{g^2 N_c}{(4\pi)^2}, & \alpha_y & = \frac{y^2 N_c}{(4\pi)^2}, & \alpha_h & = \frac{u N_f}{(4\pi)^2}, & \alpha_v & = \frac{v N_f^2}{(4\pi)^2}\ .
\end{align}
Note that this $4$D UV fixed-point of the Banks-Zaks type exists only in the presence of gauge bosons since arbitrary small parameter $\delta$ emerges measuring deviation from the asymptotic freedom boundary at $N_f/N_c=11/2$. For $\delta>0$, the asymptotic freedom is lost and the model has the perturbative asymptotically safe fixed-point instead. In contrast, the pure scalar sector leads to $U(N_f)\times U(N_f)$ linear sigma model and does not have any small parameter to define perturbative fixed-point in $4$D. A family of large charge operators in this linear sigma model in $D=4-\epsilon$, where the infrared Wilson-Fisher fixed-point still exists, was considered in \cite{Antipin:2020rdw}. As demonstrated there, even though for $N_f>\sqrt{3}$ this fixed-point is complex the method still applies for any $N_f$.

We follow \cite{Antipin:2020rdw} and consider a homogeneous ground state 
\begin{equation}
\Phi_0\left(\tau\right)=e^{2iM \tau} B\ ,
\label{eq:ansatz}
\end{equation}
where $M$ and $B$ are $N_f \times N_f$ diagonal matrices with entries $M_{ii} = -i\mu_i$ and $B_{ii} = b_i$.
The Noether charges are two diagonal and traceless $N_f \times N_f$ matrices given by
\begin{align}
{{\mathcal{Q}}_{L}}=-V{{\dot{\Phi}}_{0}}\Phi_{0}^{\dagger }= -2iVM_E{{B}^{\dagger }}B \equiv \mathcal{Q} , \qquad {\mathcal{Q}_{R}}=V\Phi_{0}^{\dagger }{{\dot{\Phi}}_{0}} = 2iVM_E{{B}^{\dagger }}B = -\mathcal{Q}\,\,.
\label{eq:qdef}
\end{align}
where $\displaystyle{V=\frac{2 \pi^{D/2}}{\Gamma(D/2)}}$ is the volume of $S^{D-1}$. The equations of motion read
\begin{equation}
    2 \mu_i^2  = u b_i^2 +v \sum_{k=1}^N b_k^2 +\frac{m^2}{2} \,.
\end{equation}
We consider a family of charge configurations labelled by $2$-parameters $J$ and $s$ as \begin{equation} \label{complete}
    \mathcal{Q}_{J,s} = \text{diag}\big(\underbrace{J,J, \dots}_s , \underbrace{-J, -J, \dots}_s , \underbrace{0, 0, \dots}_{N_f-2s} \big) \,.
\end{equation}
The above has been investigated in \cite{Antipin:2021akb} and interpolates between the charge matrices considered in \cite{Antipin:2020rdw} ($s=1)$ and \cite{Orlando:2019hte} ($s=N_f/2$).
The $M$ and $B$ matrices can be parametrized as
\begin{equation}
     \mu_i =  \begin{cases}
 \mu  & i=1,\dots,s \,,\\
-\mu & i=s+1,\dots, 2s \,,\\
 0 & i=2s+1,\dots, N_f \,,
\end{cases} \qquad  \qquad    b_i =  \begin{cases}
 b  & i=1,\dots,2s \,,\\
 0 & i=2s+1,\dots, N_f \ ,
\end{cases}
\end{equation}
yielding the following equations of motion
\begin{equation}
    J = 2 V \mu b^2  \,, \qquad \qquad 2 \mu^2 =(u + 2 s v)b^2 + \frac{m^2}{2} \ .
\end{equation}
The classical scaling dimension in $D=4$ of the corresponding lowest-lying charged operator is $ Q = 4 s J$ \cite{Antipin:2021akb}. The equations of motion can be solved to obtain the chemical potential
\begin{equation} \label{xdef}
    2 \frac{\mu}{m} =\frac{3^{{1}/{3}}+x^{{2}/{3}}}{3^{{2}/{3}}x^{{1}/{3}}} \,, \qquad  x=  \frac{72 J}{N_f^2}( \alpha_h N_f + 2  s \alpha_v )  + \sqrt{-3+ \left(\frac{72 J}{N_f^2}( \alpha_h N_f +  2  s \alpha_v) \right)^2 } \,,
\end{equation}
as well as the leading order $J \Delta_{-1}$ of the semiclassical expansion for the scaling dimension \cite{Antipin:2021akb}, reading \footnote{Here $\Delta_Q = \sum_{k=-1} \frac{\Delta_k(J \alpha_h, J \alpha_v, J \alpha_y, J \alpha_g, s)}{J^k}$.}
\begin{equation} \label{LOLS}
  J  \Delta_{-1} (J \alpha_h, J \alpha_v) = \frac{N_f^2}{72 ( \alpha_h N_f + 2 s \alpha_v )} \frac{s}{ 2 x^{\frac{4}{3}}}\left(\sqrt[3]{3} x^{8/3}-3 x^{4/3}+6 \sqrt[3]{3} x^{2/3}+2\ 3^{2/3}
   x^{2}+ 3^{5/3} \right) \,.
\end{equation}
We separate the bosonic and fermionic contributions to $\Delta_0$ as
\begin{align} \label{separ}
\Delta_0 =  \Delta_0^{(b)} -2 s N_c \Delta_0^{(f)}\,\,.
\end{align}
The dispersion relations of the scalar modes and the explicit renormalized expression for $\Delta_0^{(b)}$ can be found in \cite{Antipin:2020rdw}. The dispersion relations of the fermionic modes can be found by proceeding as in the previous section. We have $2 s N_c$ modes corresponding to the fermions carrying the flavor charges that have been fixed. Their dispersion relations read
\be
\omega_{f\pm}(\ell)=\sqrt{\left(\mu+\l_{f\pm}\right)^2+\frac{y^2 N_f^2 \left(4\mu^2-m^2\right)}{32 \pi ^2 \left(N_f \alpha _h+2 s \alpha _v\right)}}\,\,.
\ee
In addition, we have $(N_f-2 s) N_c$ modes with the dispersion relation of a free massless fermion $\omega_0 = \l_{f \pm}$ that do not contribute to $\Delta_0$. Hence, since $\Delta_0^{(f)} = \sum_{\ell=0} n_f(\ell)\omega_{f\pm}(\ell)$, we have
\begin{align}\label{delta0f}
\Delta_0^{(f)}  =& -\frac{\left(4 \mu ^2-1\right) \alpha_y N_f^2 \left(\left(12 \mu ^2+1\right) \alpha_y N_f^2-2 \left(12 \mu ^2-13\right) N_c \left(N_f \alpha _h+2 s \alpha _v\right)\right)}{32 N_c^2 \left(N_f \alpha _h+2 s \alpha _v\right){}^2} \nonumber \\ &+\sqrt{\frac{2 \left(4 \mu ^2-1\right) \alpha_y N_f^2}{N_c \left(N_f \alpha _h+2 s \alpha _v\right)}+(2 \mu +3)^2} +\sqrt{\frac{2 \left(4 \mu ^2-1\right) \alpha_y N_f^2}{N_c \left(N_f \alpha _h+2 s \alpha _v\right)}+(3-2 \mu )^2} \nonumber\\&-6 + \frac{1}{2}\sum_{l=1} \sigma^{(f)}(\ell)  \,,
&
\end{align}
where
\begin{align}
  \sigma^{(f)}(\ell) =&\,\, \frac{\left(1-4 \mu ^2\right)^2 N_f^4 \alpha_y^2}{8  \ell N_c^2 \left(N_f \alpha _h+2 s \alpha _v\right){}^2}-\frac{\left(4 \mu ^2-1\right) N_f^2 \left(4 \ell^2+4 \mu ^2+6 \ell-1\right) \alpha _y}{4 \ell N_c \left(N_f \alpha _h+2 s \alpha _v\right)} \nonumber \\ & + (\ell+1) (\ell+2) \Bigg( \sqrt{\frac{2 \left(4 \mu ^2-1\right) N_f^2 \alpha _y}{N_c \left(N_f \alpha _h+2 s \alpha _v\right)}+\left(3+ 2 \ell+2 \mu \right)^2} \nonumber \\ & +\sqrt{\frac{2 \left(4 \mu ^2-1\right) N_f^2 \alpha _y}{N_c \left(N_f \alpha _h+2 s \alpha _v\right)}+\left(3+ 2 \ell-2 \mu \right)^2} -4 \ell-6 \Bigg)  \,\,.
\end{align}
The small-charge expansion of $\Delta_0^{(f)}$ up to order $\mathcal{O}\left(Q^3 \right)$ reads
\be
\begin{split}
   \Delta_0^{(f)} =&\,\,\frac{Q \alpha _y \left(N_c-\frac{N_f^2 \alpha _y}{N_f \alpha _h+2 s \alpha _v}\right)}{2 s N_c^2}+ \frac{Q^2 \alpha _y \left(2 N_c N_f \alpha _h+4 s N_c \alpha _v+N_f^2 \left(-\alpha _y\right)\right)}{2 s^2 N_c^2 N_f^2} \nonumber \\ 
   &+ \frac{Q^3 \alpha _y \left(-2 N_c^2 \left(N_f \alpha _h+2 s \alpha _v\right){}^2-(3 \zeta (3)-1) N_c N_f^2 \alpha _y \left(N_f \alpha _h+2 s \alpha _v\right)\right)}{s^3 N_c^3 N_f^4} \\
   &+\frac{Q^3 \alpha _y\,\zeta (3) N_f^4 \alpha _y^2}{s^3 N_c^3 N_f^4}+\mathcal{O}\left( Q^4\right)\,\,.
\end{split}
\ee
Again, we give explicit results up to the $6$-loop order in App.\ref{App1}. In \cite{Antipin:2020rdw} it has been shown that the fixed-charge operators transform according to the irreducible representations $\left(\Gamma_L,\Gamma_R\right)$ of $SU(N_f)_L\times SU(N_f)_R$ where $\Gamma_{L}$ appears in the decomposition of the tensor power of the adjoint representation of $SU(N_f)_L$, denoted as $\left(\textbf{Adj}_{L}\right)^{\bigotimes Q/2}$ and  analogously $\Gamma_R \in \left(\textbf{Adj}_R\right)^{\bigotimes Q/2}$. The corresponding operators can be built as a product of blocks with simple definite transformation properties under $SU(N_f)_L\times SU(N_f)_R\times U(1)_A$ i.e. as a product of "unit-charge" blocks. These can be formally written as
\begin{align}
\Tr\Big[ \Pi_j (\tau_{j} \Phi\tau_{j}^\dagger \Phi^\dagger)^{y_j}\Big] \,,
\label{eq:ocon}
\end{align}
where $y_j>0$ is an integer and $\tau_{j}$ is an $N_f\times N_f$ matrix defined as
$\tau_{j} \equiv E_{p(j)q(j)}$ for some $p,q=1,2,...,N_f$ that depend on $j$. Here $E_{pq}$ is an $N_f\times N_f$ matrix with $1$ in the $(p,q)$ entry and "$0$" elsewhere. The fixed-charge operators are then obtained as products of these blocks with the freedom of redistributing the trace operation and changing the order of matrix products
for different $j$. We do not consider derivatives in the construction since they generally increase the scaling dimension. As explained in \cite{Antipin:2021akb}, this construction ensures that the charge configuration $\mathcal{Q}$ is a linear combination of the charge configurations $\mathcal{Q}_j$ of its building blocks, i.e. $\mathcal{Q}=\sum_{j}y_j \mathcal{Q}_j$, where
\begin{align}
\mathcal{Q}_j=\frac{1}{2}\alpha_{p(j)q(j)} \ ,
\end{align}
with ${\alpha }_{pq}$ being the nonzero weights of $\mathbf{Adj}_L$ which corresponds to the  nonzero roots of $SL(N_f,\mathbb{C})$ \footnote{These are given by  ~\cite{Hall:2015tb}
\begin{align}
{{\alpha }_{jk}}={{e}_{j}}-{{e}_{k}} \ , \qquad\text{  }j\ne k \ , \qquad\text{  }j,k=1,2,...,N_f \ ,
\label{eq:slncweight}
\end{align}
where $e_j$'s denote the standard basis elements of ${{\mathbb{C}}^{N_f}}$, i.e. ${{e}_{j}}=\{\underbrace{0,...,0}_{j-1},1,\underbrace{0,...,0}_{N_f-j}\}$.}. Note that according to our normalization, the charge configuration is related to the weight $\mathbf{w}$ appearing in the corresponding irreducible representation as $\mathcal{Q}= \mathbf{w}/2$. In general, this construction does not predict which operator has the lowest scaling dimension $\Delta_Q$. However, in the case $s=1$ one can prove that the lowest-lying operators live in the $(\Gamma_J,\Gamma_J)$ representation of $SU(N_f)_L\times SU(N_f)_R$ where $\Gamma_J$ has Dynkin label $(2J,0,0, \dots, 0, 2J)$ \cite{Antipin:2021akb}. In particular, for $s=1$, $\Delta_{Q=2}$ is the scaling dimension of the operator $\Tr[T^a \Phi T^a \Phi^\dagger ]$ which transforms in the $(\mathbf{Adj}_L,\mathbf{Adj}_R)$ representation of $SU(N_f)_L \times SU(N_f)_R$ \footnote{The adjoint representation of $SU(N_f)$ has Dynkin label $(1,0,0, \dots, 0, 1)$.}. Since the scaling dimension of this operator has been computed to two loops in \cite{Antipin:2014mga}, we can combine this information with our results to obtain $\Delta_{Q,s=1}$ to the same order. Moreover, we note that since the scalars are not charged under the gauge group then the contribution of the gauge coupling to $\Delta_{Q,s=1}$ starts (at least) at next-to-next-to-leading order (NNLO) in the semiclassical expansion (i.e. with $\Delta_1$). Therefore, there is no contribution at $1$-loop and the $2$-loop contribution is simply linear in $Q$. Hence it can be fixed by matching with $\Delta_{\Tr[T^a \Phi T^a \Phi^\dagger]}$. We find
\begin{align} \label{toloop}
\Delta_{Q,s=1}^{(2-\text{loop})}&=Q\left(\frac{d-2}{2}\right) + \frac{(Q-2) Q \alpha _h}{N_f}+\frac{2 (Q-1) Q \alpha _v}{N_f^2}+Q \alpha _y -Q \left[2 \left(\frac{3}{N_f^2}-\frac{4}{N_f}-1\right) \alpha _h^2 \right. \nonumber \\ & \left. +8 \left(\frac{2}{N_f^3}-\frac{3}{N_f^2}\right) \alpha _h \alpha _v+2 \left(\frac{1}{N_f^4}-\frac{3}{N_f^2}\right) \alpha _v^2-\frac{4 \alpha _h \alpha _y}{N_f}-\frac{4 \alpha _v \alpha _y}{N_f^2}  \right. \nonumber \\ & \left.+z \left(\frac{3}{2}+\frac{2}{N_f}\right) \alpha _y^2-\frac{5}{2} \left(1-\frac{z^2}{N_f^2}\right)\alpha _g \alpha _y\right] + Q^2 \left[2 \left(\frac{1}{N_f^2}-\frac{2}{N_f}\right) \alpha _h^2  \right. \nonumber \\ & \left.+8 \left(\frac{3}{N_f^3}-\frac{2}{N_f^2}\right) \alpha _h \alpha _v+4 \left(\frac{3}{N_f^4}-\frac{1}{N_f^2}\right) \alpha _v^2-\frac{2 \alpha _h \alpha _y}{N_f}-\frac{4 \alpha _v \alpha _y}{N_f^2}+\frac{z \alpha _y^2}{N_f} \right]  \nonumber \\ & -\frac{2 Q^3}{N_f^4}\left(N_f \alpha _h+2 \alpha _v\right){}^2 \,,
\end{align}
which include the contribution of all the couplings of the theory. The contribution from the quartic couplings has been previously derived in \cite{Antipin:2020rdw}.

\subsection{$\Delta_Q$ in the Veneziano limit and the large-charge expansion}

It is interesting to study the Veneziano limit \eqref{Veneziano} of our results. In fact, this is the limit where the perturbative asymptotically safe fixed point is realized. Moreover, as we shall see the dynamics simplify considerably due to factorization. We take the Veneziano limit by keeping $Q$ and $1/\delta$ large but parametrically smaller than $N_f$, $N_C$ , i.e. we consider $N_f, N_c \gg Q, 1/\delta \gg 1$. Moreover, for the moment we consider that $s$ does not scale with $N_f$. In this case, from eqs.\eqref{LOLS}, \eqref{delta0f}, and the expression of $\Delta_0^{(b)}$ obtained in \cite{Antipin:2021akb}, we obtain 
in the Veneziano limit 
\be \label{Venicebeach}
\Delta_Q= \frac{Q}{4s}\Delta_{-1}+\Delta_0 + \mathcal{O}\left(\frac{4s}{Q}\Delta_1\right)=Q \bigg[ \underbrace{1}_{\Delta_{-1}}\underbrace{-4  \alpha _h}_{\Delta_0^{(b)}}
  \underbrace{+\frac{ z \alpha _y^2}{\alpha _h}- \alpha_y}_{-2 s N_c \Delta_0^{(f)}}\bigg]  + \mathcal{O}\left(\frac{4s}{Q}\Delta_1\right)
  \,\, .
\ee
 The expression \eqref{Venicebeach} is remarkably simple: in the full $\Delta_{-1}$ and $\Delta_0$ survives only the term linear in $Q$ of their small-charge expansion. Moreover, the $s$-dependence disappears completely. We conjecture this to be true at the non-perturbative level, i.e. 
\be \label{free}
\Delta_Q = Q \Delta_{Q=1} = Q \Delta_{\Phi} \,,
\ee
and we interpret the above as the occurrence of a \emph{generalized free field theory phase} in the large-charge sector of the theory. This should be seen in contrast with the superfluid phase realized in generic CFT, which leads to eq.\eqref{largecharge} in the large-charge regime as opposed to $\Delta_Q \approx Q$ \footnote{Even if in taking the Veneziano limit we did not treat the charge as the largest parameter of our theory, one may still consider the charge the largest parameter of the ultraviolet CFT defined in such a limit.}. The latter behaviour is indeed realized in free field theories. As we shall see below, in this limit there is an operator corresponding to $Q=1$, which is simply $\Phi$. 

Note that 
eq.\eqref{free} can be seen as a consequence of large $N_{f}$ factorization. In fact, we saw that the fixed-charge operators live in tensor powers of the adjoint representation. At the same time, complete factorization in the $(\mathbf{Adj}_L,\mathbf{Adj}_R)$ channel has been previously discussed in \cite{Antipin:2014mga, Heemskerk:2009pn} yielding the OPE
\be
\Phi(x_1)\times \Phi^\dagger(x_2) = \frac{1}{x_{12}^{2 \Delta_\Phi}}+\sum_{n,l}\frac{c_{n,l}^A}{x_{12}^{2n+l}}\mathcal{O}_{n,l}^A \,,
\ee
where we have only the contribution of double-trace operators
\be 
O_{n,l}^A =\big(\mathcal{O}^A \overset{\leftrightarrow}{\partial}_{\mu_1} \dots \overset{\leftrightarrow}{\partial}_{\mu_l} (\overset{\leftrightarrow}{\partial}_{\nu} \overset{\leftrightarrow}{\partial^\nu})^n \mathcal{O}^A -\text{Traces} \big)\,.
\ee 
The scaling dimensions of $\mathcal{O}_{n,l}^A$ is $\Delta_{n,l} = 2 \Delta_\Phi +2 n +l$ where $l$ denotes its spin. The first operator is $O_{n,l}^A = \Tr[T^a \Phi T^a \Phi^\dagger] = \frac12 \Tr[\Phi] \Tr[\Phi^\dagger]+ \mathcal{O}(\frac{1}{N_f})$, where we used the well-known identity $T_{ij}^a T_{kl}^a = \frac12 \delta_{il} \delta_{jk} -\frac{1}{2 N_f} \delta_{ij} \delta_{kl}$ for the $SU(N_f)$ generators.
In particular, the latter observation explains why we have $\Delta_\Phi = \Delta_{Q=1}$ in the Veneziano limit. 

From the point of view of the operator construction, we observe that large-$N_f$ factorization occurs at the level of the single trace operators, i.e. implies that the building blocks $(\tau_{j} \Phi\tau_{j}^\dagger \Phi^\dagger)^{y_j}$ introduced in eq.\eqref{eq:ocon} appears inside the same trace. 

Finally, in the Veneziano limit eq.\eqref{toloop} reduces to 
\be \label{venicebeach}
\Delta_{Q,s=1}^{(2-\text{loop})} = Q\left(\frac{d-2}{2}\right) +Q \left(\alpha _y+\frac{5 \alpha _g \alpha _y}{2}+2 \alpha _h^2-\frac{3 z \alpha _y^2}{2}\right)\,.
\ee
For $Q=1$, eq.\eqref{venicebeach} agrees with the known $2$-loop scaling dimension of $\Phi$ in the Veneziano limit \cite{Antipin:2014mga}.

In order to realize the superfluid phase characterized by eq.\eqref{largecharge}, we should let the charge scale with $N_f$ in the Veneziano limit. In particular, from eq.\eqref{xdef}, we see that the natural 't Hooft coupling is not $J \delta$ but rather 
\be
\mathcal{J}=\frac{2 J}{N_f^2}\left(N_f \alpha _h+2 s \alpha _v\right)\,.
\ee
We, therefore, take the Veneziano limit by keeping $\mathcal{J}$ fixed and large. In other words, we are interested in the regime where the charge is the dominant large parameter of our CFT. This is defined by the condition $\mathcal{J}\gg 1$, that is
\be
 J \gg \frac{N_f}{\delta}\,\,.
\ee
For simplicity, we will restrict ourselves to the case $s=s_{\text{max}}=N_f/2$, previously considered in \cite{Orlando:2019hte}. $\Delta_{-1}$ can be trivially expanded for large $\mathcal{J}$, leading to eq.(3.24) of \cite{Orlando:2019hte}. To obtain the large $\mathcal{J}$ expansion of $\Delta_0$ we first take the Veneziano limit of eq.\eqref{separ} which leads to $\frac{\Delta_0}{N_f^2} = \Tilde{\Delta}_0(\alpha_h, \alpha_v, \alpha_y, z, \mu)$. Next, we substitute the couplings with their fixed point value \eqref{LSFP}, use that $z=\delta+11/2$, and expand $\Tilde{\Delta}_0$ around $\delta=0$ to the leading order. Finally, we split the sum and use the Euler-Maclaurin formula as explained in the previous section, obtaining 
\begin{align}
\frac{\Delta_Q}{N_f^2}=&\frac{361}{4 \left(\sqrt{23}-\sqrt{6 \sqrt{23}+20}+1\right)^2}\frac{\mathcal{J}^2}{\delta^2 J^2}\Delta_Q=  \frac{\mathcal{J}^{4/3}}{\delta}  \Bigg[\frac{57}{88}\left(\sqrt{23}+\sqrt{46 \sqrt{23}+189}+12\right)\, \nonumber \\  
\nonumber& -3.3777 (1) \delta + \mathcal{O} (\delta^2) \Bigg] + \frac{\mathcal{J}^{2/3}}{\delta}  \Bigg[\frac{19}{176} \left(\sqrt{23}+\sqrt{46 \sqrt{23}+189}+12\right) \\
&+4.5881 (1)\delta + \mathcal{O} (\delta^2) \Bigg]   + \mathcal{O}\left(\mathcal{J}^0\right) \,,
\end{align}
where the digits in brackets denote the numerical error owing to the numerical evaluation of the integrals in the Euler-Maclaurin formula. The result above matches with the general structure of eq.\eqref{largecharge} stemming from the superfluid EFT. Interestingly, at the considered order, the $\log \mu$ terms arising in the procedure cancel between scalars and fermions illustrating the interplay of different kinds of matter fields in realizing conformal dynamics. Note that in the case of the Wilson-Fisher fixed-point as the one in the previous section, these logarithms do not cancel but instead are crucial for obtaining the form \eqref{largecharge} for non-integer $D$. Moreover, there is no  universal $Q^0 \log Q$ term. The latter is a prediction of the superfluid effective theory describing the large-charge sector of generic $U(1)$-invariant CFT \cite{Cuomo:2020rgt}. This is an effective field theory for the superfluid phonon stemming from the combined spontaneous breaking of external and internal symmetries in states with finite charge. In fact, while this mode is present in our spectrum \cite{Antipin:2021akb}, its contribution is subleading in the Veneziano limit. The order $\mu^0$ term in the large $\mu$ expansion of $\Delta_0^{(b)}$ reads
\begin{align}
& \frac{1}{32} \left(7 \sqrt{23}+\sqrt{326 \sqrt{23}+1479}+2\right) + \frac12 \sum_{\ell=1} \Bigg\{\frac{N_f^2}{4 \ell \left(\alpha _h+\alpha _v\right){}^2} \left(\left(5-8 \ell^2 (\ell+1) (\ell+2)\right) \alpha _h^2 \right. \nonumber \\ & \left. -4 \left(4 \ell^4+12 \ell^3+9 \ell^2+\ell-1\right) \alpha _h \alpha _v+\left(-8 \ell^4-24 \ell^3-20 \ell^2-4 l+1\right) \alpha _v^2\right) \nonumber \\ & \frac{2 \left(2 \ell^2+2 \ell+3\right) \alpha _h \alpha _v+4 \left(\ell^2+\ell+1\right) \alpha _v^2}{4 \ell \left(\alpha _h+\alpha _v\right){}^2}+\frac{\sqrt{\ell (\ell+2)} (\ell+1)^2}{ \sqrt{3}} \Bigg\} \,,
\end{align}
where the last term can be identified with the contribution of the superfluid phonon with the characteristic speed $\frac{1}{\sqrt{d-1}}=\frac{1}{\sqrt{3}}$ \cite{Hellerman:2015nra}.

\section{Conclusions and  future directions}

\label{conclusions}


In this work we generalized the semiclassical method of \cite{Badel:2019oxl} to compute the contributions to the anomalous dimensions of fixed-charge scalar operators from Yukawa  interactions. The models we studied were  the  NJLY and  an asymptotically safe gauge-Yukawa model in four dimensions. In all cases, we have chosen as a vacuum a superfluid phase with homogeneous charge density so that the leading contribution of fermionic fields starts at the one-loop level with the $\Delta_0$ term. Our main results for all the models include:
\begin{enumerate}
\item Expansion of both $\Delta_{-1}$ and $\Delta_0$ in the \emph{small} 't Hooft coupling limit. In Appendix \ref{riscomp} we have collected explicit perturbative results up to the $6$-loop level stimulating future comparisons with other computational methods.
\item Expansion of both $\Delta_{-1}$ and $\Delta_0$ in the \emph{large} 't Hooft coupling limit. We have derived EFT predictions inviting comparisons with future Monte-Carlo and lattice studies.
\item Beyond $\Delta_{-1}$ and $\Delta_0$ terms, we have \emph{boosted} perturbative expansion by matching with the known $2$-loop perturbative result for $Q=1$ operator, predicting the full $2$-loop result for operators with any $Q$.
\end{enumerate}
Main model-specific results can be summarized as:
\begin{itemize}
\item For the NJLY model, we gave evidence for emergent supersymmetry in the critical model by looking at the dispersion relations of the fermions for $N_f = 1/2$. Moreover, by exploiting that the $\phi^2$ operator is a descendant of $\phi$ at the supersymmetric fixed point, we obtained $\Delta_Q$ to order $\cO\left(\e^3 \right)$ in the Wess-Zumino model.

\item For the asymptotically safe gauge-Yukawa model, we demonstrated factorization in the Veneziano limit and the emergence of a generalized Gaussian CFT phase of the theory when the total charge does not scale with $N_f$, $N_c$. In the opposite case, the theory realizes a superfluid-like phase for which we observed the absence of the expected universal $Q^0 \log Q$ contribution to $\Delta_Q$, since this is suppressed in the considered limit.

\end{itemize}

Several directions could be followed since the formalism above can be applied to general theories with other types of global symmetries.  The prominent theoretical direction may be the evaluation of the anomalous dimension of the large-charge operators in $\mathcal N=4$ SYM  which will shed some light on the structure of the AdS/CFT correspondence in this limit  as well as the phase diagram of the theory 
\cite{Yamada:2006rx,Hollowood:2008gp}.

For applications  of the large charge expansion to two-dimensional models, an interesting class of $\s$-models are the $\l$-deformations \cite{Sfetsos:2013wia} of WZW-models 
where a class of chiral chain operators have been calculated \cite{Georgiou:2019jcf} at large $k-$level and classical solutions have been obtained in \cite{Katsinis:2021nfu,Hernandez:2022dmf}. Following the general lines of \cite{Araujo:2021mdm} we may use the large charge expansion to study the interplay between large charge and large $k$. 

It would be also interesting to extend the method to the study of fermionic operators. For the free theory, the anomalous dimensions of the charged fermions were obtained in \cite{Komargodski:2021zzy}. Nevertheless, it remains an open question to apply the method to fermionic operators in interacting theories such as the Gross-Neveu model \cite{Gross:1974jv} which may be specified by fermion-boson dualities in two dimensions \cite{Giombi:2014xxa}. Also, since WZW-model has fermionic representation \cite{DiFran} with $k=1$ it is possible to extract information about fermions in the large charge limit since  large $k$ calculations are invalid.

\section*{Acknowledgments}

The  work of P. Panopoulos was supported by the Croatian Science Foundation Project "New Geometries for Gravity and Spacetime" (IP-2018-01-7615). We wish to thank Simeon Hellerman, Alexander Monin and Francesco Sannino for valuable discussions.

\appendix

\section{Scalar and spinor  fields on $S^{D}$}

\label{sphere}

We give some details about the Laplacian of scalar  and spinor fields  defined on $S^{D}$ sphere. In this appendix we restore the radius of $S^D$ that we denote as $R$. The conformally coupled action for a real scalar field $\phi$ on a curved background with Ricci scalar $\mathcal R$  is written as
\small
\be
\begin{split}
S=&\frac{1}{2}\int d^D x\sqrt{-g}\left((\del_{\m}\phi)^2+\frac{D-1}{4D}\mathcal R\phi^2\right)
=\frac{1}{2}\int d^Dx\sqrt{-g}\,\phi\left(-\nabla^2_{S^D }+\frac{D-1}{4D}\mathcal R\right)\phi \,.
\end{split}
\ee
\normalsize

Recall that the Laplacian $\D$, on curved background $\mathcal M$ is defined by
\be
\Delta \phi=-\frac{1}{\sqrt{g}}\del_{\m}\left(\sqrt{-g}g^{\m\n}\del_{\n}\phi\right)=-\nabla^2_{\mathcal M}\phi \,,
\ee
where $\nabla^2\equiv\nabla_{\m}\nabla^{\m}$ and as usual $\nabla_{\m}U^{\n}=\del_{\m}U^{\n}+\G^{\n}_{\m\rho}U^{\rho}$. The eigenvectors of the scalar Laplacian are spherical harmonics $Y_{\ell}$ labelled by angular momentum quantum numbers, $\ell\in \mathbb{Z}\geq0$ with 
\be
\label{eigenscalar}
-\nabla^2_{S^D}\,Y_{\ell}=\frac{1}{R^2}\ell(\ell +D-1)Y_{\ell} \,,
\ee
and degeneracy 
\be
n_b(\ell)=\frac{(2\ell+D-1)\G(\ell+D-1)}{\G(D)\G(\ell+1)}\,\,.
\ee
Finally the $S^D$ Laplacian  acting on spinor fields, gives the following eigenvalues and degeneracy respectively
\be
\l_{f\pm}(\ell)=\pm\left(\ell+\frac{D}{2}\right),\qquad 
 n_f(\ell)=\frac{4\,\G(\ell+D)}{\G(D)\G(\ell+1)}\,\,.
\ee

\section{Integrals appearing in the large-charge expansion of $\Delta_0$ in the NJLY model} 

\label{App1}
Here, for reader's convenience, we give the two integrals  appearing in eq.\eqref{high}. The first one is  
\be
\begin{split}
  \Sigma_1^{\text{reg}}(k) = \frac{1}{8 k} \Big(&-8 k^4-8 k^2+4 \sqrt{k^2
  -\sqrt{4 k^2+9}+3} k^3\\
  &+4 \sqrt{k^2+\sqrt{4 k^2+9}+3} k^3+5\Big) \nonumber  -\frac{5}{(8 k) \left(\frac{5 k^2}{8}+1\right)} \,,
  \end{split}
  \ee
and the second one reads
\be
\begin{split}
   \Sigma_3^{\text{reg}}(k) = \frac{1}{2} & \Bigg[ \frac{\left(2-\frac{4}{\sqrt{4 k^2+9}}\right) k^2}{\sqrt{k^2-\sqrt{4 k^2+9}+3}}+\frac{\left(\frac{20 k^2+27}{\left(4 k^2+9\right)^{3/2}}-1\right) k^2}{2 \sqrt{k^2-\sqrt{4 k^2+9}+3}}+\frac{\left(\frac{4}{\sqrt{4 k^2+9}}+2\right) k^2}{\sqrt{k^2+\sqrt{4 k^2+9}+3}} \nonumber \\ &  +\sqrt{k^2-\sqrt{4 k^2+9}+3}+\sqrt{k^2+\sqrt{4 k^2+9}+3}-\frac{\left(2-\frac{4}{\sqrt{4 k^2+9}}\right)^2 k^4}{8 \left(k^2-\sqrt{4 k^2+9}+3\right)^{3/2}} \nonumber \\ & -\frac{\left(3 \left(18 \sqrt{4 k^2+9}+65\right) k^2+54 \left(\sqrt{4 k^2+9}+3\right)+\left(8 \sqrt{4 k^2+9}+52\right) k^4\right) k^2}{2 \left(4 k^2+9\right)^{3/2} \left(k^2+\sqrt{4 k^2+9}+3\right)^{3/2}}\Bigg]\\
   &-2 k-\frac{5}{4 k}  +\frac{5}{(4 k) \left(\frac{5 k^2}{4}+1\right)}\,\,\,.
\end{split}
\ee

\section{Explicit results for $\Delta_Q$} \label{riscomp}

In this appendix we provide explicit results for $\Delta_Q$ for all the models considered in this work. The expressions include the contributions stemming from the small 't Hooft coupling expansion of both $\Delta_{-1}$ and $\Delta_0$. We write the usual loop-expansion for $\Delta_Q$ as
\be
\Delta_Q =Q\left(\frac{d-2}{2}\right) + \sum_{l=1} P_Q^{(l-\text{loop})} \,,
\ee
where $P_Q^{(l-\text{loop})}$ is a polynomial of degree $l+1$ in $Q$, i.e.
\be
P_Q^{(l-\text{loop})} = \sum_{k=0}^{l} C_{kl} Q^{l+1-k} \,.
\ee
By comparing with eq.\eqref{PT}, we see that the $C_{kl}$ stems from the small-charge expansion of $\Delta_{k-1}$ i.e. from the $k$-th order of the semiclassical expansion. For every model, we list $C_{0l}$ and $C_{1l}$ up to $l=6$.
\subsection{NJLY model}
For the NJLY model we obtain
\small
\begin{align} \label{cicci}
C_{01}&= \frac{\lambda }{3} \,,\quad C_{02}= -\frac{2}{9} \lambda ^2 \,,\quad  C_{03}=\frac{8 \lambda ^3}{27} \,, \nonumber \\  C_{04}&=-\frac{14}{27}  \lambda ^4\,,\quad   C_{05}= \frac{256 \lambda ^5}{243}\,,\quad   C_{06}= -\frac{572 }{243}\lambda ^6 \,,
\end{align}
and
\begin{align}
C_{11}&= \frac{g^2 N_f}{8 \pi ^2}-\frac{\lambda }{3}\,,\quad C_{12}=\frac{g^4 N_f}{32 \pi ^4}-\frac{g^2 \l N_f}{12 \pi ^2}+\frac{2 \lambda ^2}{9} \,, \nonumber \\   C_{13}&=-\frac{g^6 \zeta (3) N_f}{64 \pi ^6}+\frac{g^4 \lambda  (3 \zeta (3)-1) N_f}{48 \pi ^4}+\frac{g^2 \lambda ^2 N_f}{18 \pi ^2}+\frac{2}{27} \lambda ^3 (16 \zeta (3)-17) \,, \nonumber \\  C_{14}&=\frac{5 g^8 \zeta (5) N_f}{1024 \pi ^8}+\frac{g^6 \lambda  (6 \zeta (3)-5 \zeta (5)) N_f}{192 \pi ^6}+\frac{g^4 \lambda ^2 (4-15 \zeta (3)) N_f}{144 \pi ^4}\nonumber \\ &-\frac{g^2 \lambda ^3 (\zeta (3)+4) N_f}{54 \pi ^2} -\frac{2}{81} \lambda ^4 (77 \zeta (3)+80 \zeta (5)-142)\,, \nonumber \\  C_{15}&=-\frac{7 g^{10} \zeta (7) N_f}{4096 \pi ^{10}}+\frac{5 g^8 \lambda  (7 \zeta (7)-8 \zeta (5)) N_f}{3072 \pi ^8}+ \frac{g^6 \lambda ^2 (5 \zeta (5)-7 \zeta (3)) N_f}{96 \pi ^6} \nonumber \\ &+ \frac{g^4 \lambda ^3 (96 \zeta (3)+10 \zeta (5)-21) N_f}{432 \pi ^4} +\frac{g^2 \lambda ^4 (10 \zeta (3)+21) N_f}{162 \pi ^2}  \nonumber \\ & + \frac{2}{243} \lambda ^5 (476 \zeta (3)+480 \zeta (5)+448 \zeta (7)-1179)\,,  \nonumber \\  C_{16}&= \frac{21 g^{12} \zeta (9) N_f}{32768 \pi ^{12}} +\frac{7 g^{10} \lambda  (10 \zeta (7)-9 \zeta (9)) N_f}{12288 \pi ^{10}} +\frac{5 g^8 \lambda ^2 (64 \zeta (5)-49 \zeta (7)) N_f}{9216 \pi ^8}  \nonumber \\  & + \frac{5 g^6 \lambda ^3 (64 \zeta (3)-40 \zeta (5)-7 \zeta (7)) N_f}{1728 \pi ^6} -\frac{g^4 \lambda ^4 (693 \zeta (3)+115 \zeta (5)-128) N_f}{1296 \pi ^4}  \nonumber \\  & -\frac{g^2 \lambda ^5 (89 \zeta (3)+\zeta (5)+128) N_f}{486 \pi ^2} - \frac{2}{729}  \lambda ^6 (3294 \zeta (3)+3202 \zeta (5)+3360 \zeta (7)+2688 \zeta (9)-10063) \,.
\end{align}

\subsection{An asymptotically safe model in $D=4$}
\normalsize
For the asymptotically safe model considered in Sec.\ref{Litim-Sannino} we find the following coefficients
\small
\begin{align}
C_{01}&=  \frac{N_f \alpha _h+2 s \alpha _v}{s N_f^2} \,,\quad C_{02}= -\frac{2 \left(N_f \alpha _h+2 s \alpha _v\right){}^2}{s^2 N_f^4}\,,\quad  C_{03}=\frac{8 \left(N_f \alpha _h+2 s \alpha _v\right){}^3}{s^3 N_f^6} \,, \nonumber \\  C_{04}&=-\frac{42 \left(N_f \alpha _h+2 s \alpha _v\right){}^4}{s^4 N_f^8}\,,\quad   C_{05}= \frac{256 \left(N_f \alpha _h+2 s \alpha _v\right){}^5}{s^5 N_f^{10}}\,,\quad   C_{06}=-\frac{1716 \left(N_f \alpha _h+2 s \alpha _v\right){}^6}{s^6 N_f^{12}} \,,
\end{align}
and
\begin{align}
C_{11}&= -\frac{2 s \alpha _h}{N_f}-\frac{2 \alpha _v}{N_f^2}+\alpha _y \,, \nonumber \\  C_{12} &= \frac{2 \alpha _h^2 \left(s-2 N_f\right)}{s N_f^2} +\frac{8 \alpha _h \alpha _v \left(-2 s N_f+2 s^2+1\right)}{s N_f^3} -\frac{4 \left(N_f^2-3\right) \alpha _v^2}{N_f^4}-\frac{2 \alpha _h \alpha _y}{s N_f}-\frac{4 \alpha _v \alpha _y}{N_f^2}  +\frac{\alpha _y^2}{s N_c} \nonumber \\  C_{13} &=\frac{8 \left(2 \zeta (3) \left(N_f+9 s\right)+N_f-8 s\right)}{s^2 N_f^3} \a_h^3 + \frac{8  \left(6 N_f (2 s \zeta (3)+s)+s^2 (30 \zeta (3)-28)+12 \zeta (3)-13\right)}{s^2 N_f^4} \alpha _h^2 \alpha _v \nonumber \\ &+ \frac{4}{s^2 N_f^2}  \alpha _y \alpha _h^2 + \frac{8 \left(N_f^2+8 N_f s(3  \zeta (3)+1)-4 \left(6 s^2-9 \zeta (3)+11\right)\right)}{s N_f^5}  \alpha _v^2  \alpha _h + \frac{2 (3 \zeta (3)-1)}{s^2 N_c N_f}  \alpha _y^2 \alpha_h  \nonumber \\ & +\frac{16 }{s N_f^3}\alpha _v \alpha _y \alpha_h +\frac{16  \left(2 \zeta (3) \left(N_f^2+7\right)+N_f^2-18\right)}{N_f^6} \alpha _v^3 +\frac{16 }{N_f^4}\alpha _v^2 \alpha _y+\frac{4 (3 \zeta (3)-1)}{s N_c N_f^2}\alpha _v \alpha _y^2-\frac{2 \zeta (3)}{s^2 N_c^2}  \alpha _y^3 \nonumber \\  C_{14}  &= -\frac{4 \left(2 (12 \zeta (3)+5 \zeta (5)+4) N_f+s (197 \zeta (3)+265 \zeta (5)-138)\right)}{s^3 N_f^4} \alpha_h^4 -\frac{4 (\zeta (3)+4)}{s^3 N_f^3} \alpha_h^3 \alpha_y \nonumber \\ & -\frac{16 \left(4 s (12 \zeta (3)+5 \zeta (5)+4) N_f+2 s^2 (93 \zeta (3)+90 \zeta (5)-89)+23 \zeta (3)+40 \zeta (5)-55\right)}{s^3 N_f^5} \alpha_h^3 \alpha_v \nonumber \\ & -\frac{8 \left(8 s (36 \zeta (3)+15 \zeta (5)+10) N_f+4 N_f^2+6 s^2 (58 \zeta (3)+50 \zeta (5)-101)+252 \zeta (3)+360 \zeta (5)-581\right)}{s^2 N_f^6} \alpha_h^2 \alpha_v^2  \nonumber \\ & -\frac{24 (\zeta (3)+4) }{s^2 N_f^4} \alpha _v \alpha _y \alpha_h^2  + \frac{8-30 \zeta (3)}{s^3 N_c N_f^2}\alpha _y^2 \alpha_h^2 -\frac{48 (\zeta (3)+4)}{s N_f^5}  \alpha _v^2  \alpha _y \alpha _h +\frac{8 (4-15 \zeta (3)) }{s^2 N_c N_f^3} \alpha _y^2 \alpha _h \alpha _v \nonumber \\ &  + \frac{64 \left(-N_f \left((3 \zeta (3)+2) N_f+4 s (9 \zeta (3)+5 \zeta (5)+2)\right)+42 s^2-56 \zeta (3)-70 \zeta (5)+127\right)}{s N_f^7} \alpha _v^3 \alpha _h  \nonumber \\ & +\frac{2 (6 \zeta (3)-5 \zeta (5))}{s^3 N_c^2 N_f} \alpha _y^3 \alpha _h  -\frac{32 \left((12 \zeta (3)+5 \zeta (5)+4) N_f^2+65 \zeta (3)+75 \zeta (5)-146\right)}{N_f^8} \a_v^4 \nonumber \\ & -\frac{32 (\zeta (3)+4) }{N_f^6} \alpha _v^3 \alpha _y + \frac{8 (4-15 \zeta (3)) }{s N_c N_f^4} \alpha _y^2 \alpha _v^2 +\frac{4 (6 \zeta (3)-5 \zeta (5)) }{s^2 N_c^2 N_f^2} \alpha _y^3 \alpha_v +\frac{5 \zeta (5) }{2 s^3 N_c^3} \alpha _y^4 \,.
\end{align}
\normalsize
For $C_{15}$ and $C_{16}$ we introduce the following notation: we denote as $A_{ijk}$ and $B_{ijk}$ the coefficients of $\alpha_h^i \a_v^j \a_y^k$ in $C_{15}$ and $C_{16}$, respectively. We obtain
\small
\begin{align}
    A_{500} &= \frac{8 \left((84 \zeta (3)+40 \zeta (5)+14 \zeta (7)+21) N_f+s (658 \zeta (3)+980 \zeta (5)+1050 \zeta (7)-579)\right)}{s^4 N_f^5} \nonumber \\ A_{410} & = \frac{16}{s^4 N_f^6} \bigg( 5 s (84 \zeta (3)+40 \zeta (5)+14 \zeta (7)+21) N_f+s^2 (1906 \zeta (3)+35 (68 \zeta (5)+53 \zeta (7)-58)) \nonumber \\ &+125 \zeta (3)+180 \zeta (5)+280 \zeta (7)-464\bigg) \nonumber \\ A_{401} & = \frac{40 \zeta (3)+84}{s^4 N_f^4} \qquad \qquad  A_{302}  = \frac{2 (96 \zeta (3)+10 \zeta (5)-21)}{s^4 N_c N_f^3} \qquad \qquad A_{311} = \frac{320 \zeta (3)+672}{s^3 N_f^5} \nonumber \\ A_{320} & = \frac{8}{s^3 N_f^7} \bigg(112 \zeta (3) \left(30 s N_f+66 s^2+17\right)+40 \zeta (5) \left(40 s N_f+198 s^2+67\right) \nonumber \\ &+560 \zeta (7) \left(s N_f+9 s^2+6\right)+756 s N_f+21 N_f^2-10588 s^2-6706\bigg) \nonumber \\ A_{230} &= \frac{16}{s^2 N_f^8} \bigg(4 s (756 \zeta (3)+400 \zeta (5)+140 \zeta (7)+147) N_f+21 (4 \zeta (3)+3) N_f^2 \nonumber \\ &+4 s^2 (596 \zeta (3)+590 \zeta (5)+350 \zeta (7)-1517)+2728 \zeta (3)+3540 \zeta (5)+3920 \zeta (7)-9050\bigg) \nonumber \\ A_{203} &=   \frac{60 \zeta (5)-84 \zeta (3)}{s^4 N_c^2 N_f^2} \qquad \qquad  A_{221} = \frac{96 (10 \zeta (3)+21)}{s^2 N_f^6} \qquad \qquad  A_{212} = \frac{12 (96 \zeta (3)+10 \zeta (5)-21)}{s^3 N_c N_f^4} \nonumber \\ A_{140} &= \frac{32 }{s N_f^9} \bigg(8 s (126 \zeta (3)+80 \zeta (5)+35 \zeta (7)+21) N_f+(168 \zeta (3)+40 \zeta (5)+63) N_f^2-1280 s^2  \nonumber \\&+1708 \zeta (3)+2040 \zeta (5)+2100 \zeta (7)-5402\bigg) \nonumber \\ A_{104} &= \frac{5 (7 \zeta (7)-8 \zeta (5))}{2 s^4 N_c^3 N_f} \qquad \qquad A_{131} = \frac{128 (10 \zeta (3)+21)}{s N_f^7} \qquad \qquad A_{113} = \frac{48 (5 \zeta (5)-7 \zeta (3))}{s^3 N_c^2 N_f^3} \nonumber \\ A_{122} &= \frac{24 (96 \zeta (3)+10 \zeta (5)-21)}{s^2 N_c N_f^5} \qquad \qquad A_{005} =-\frac{7 \zeta (7)}{2 s^4 N_c^4}  \nonumber \\ A_{050} &= \frac{64 \left((84 \zeta (3)+40 \zeta (5)+14 \zeta (7)+21) N_f^2+392 \zeta (3)+440 \zeta (5)+434 \zeta (7)-1200\right)}{N_f^{10}} \nonumber \\ A_{041} &= \frac{64 (10 \zeta (3)+21)}{N_f^8}  \qquad \qquad A_{014} = \frac{5 (7 \zeta (7)-8 \zeta (5))}{s^3 N_c^3 N_f^2}  \qquad \qquad A_{023} = \frac{48 (5 \zeta (5)-7 \zeta (3))}{s^2 N_c^2 N_f^4} \nonumber \\ A_{032} &= \frac{16 (96 \zeta (3)+10 \zeta (5)-21)}{s N_c N_f^6} \,,
\end{align}
\normalsize
and 
\small
\begin{align}
    B_{600} =& -\frac{4}{s^5 N_f^6} \bigg(4 (320 \zeta (3)+160 \zeta (5)+70 \zeta (7)+21 \zeta (9)+64) N_f \nonumber \\ & +s (9694 \zeta (3)+15042 \zeta (5)+19880 \zeta (7)+17850 \zeta (9)-9935)\bigg) \nonumber \\  B_{510} =& -\frac{8}{s^5 N_f^7} \bigg(24 s (320 \zeta (3)+160 \zeta (5)+70 \zeta (7)+21 \zeta (9)+64) N_f+4 s^2 (9478 \zeta (3)+12950 \zeta (5) \nonumber \\ &+13965 \zeta (7)+9450 \zeta (9)-11043)+1486 \zeta (3)+1946 \zeta (5)+3360 \zeta (7)+4032 \zeta (9)-8103\bigg) \nonumber \\  B_{501} =& -\frac{4 (89 \zeta (3)+\zeta (5)+128)}{s^5 N_f^5}   \qquad \qquad B_{402} = \frac{-1386 \zeta (3)-230 \zeta (5)+256}{s^5 N_c N_f^4} \nonumber \\  B_{420} =& -\frac{16 }{s^4 N_f^8}\bigg(4 s (4800 \zeta (3)+2400 \zeta (5)+1050 \zeta (7)+315 \zeta (9)+896) N_f+64 N_f^2+7416 \zeta (3)+9760 \zeta (5)  \nonumber \\  &+14840 \zeta (7)+15120 \zeta (9)-37059  +s^2 (55698 \zeta (3)+67870 \zeta (5)+60620 \zeta (7)+33390 \zeta (9)-78235)\bigg)  \nonumber \\  B_{411} =& -\frac{40 (89 \zeta (3)+\zeta (5)+128)}{s^4 N_f^6} \qquad \qquad B_{303} = \frac{10 (64 \zeta (3)-40 \zeta (5)-7 \zeta (7))}{s^5 N_c^2 N_f^3} \nonumber \\  B_{330} =& -\frac{64}{s^3 N_f^9} \bigg(8 s (1520 \zeta (3)+800 \zeta (5)+350 \zeta (7)+105 \zeta (9)+256) N_f+32 (5 \zeta (3)+4) N_f^2  \nonumber \\ &+4 s^2 (4557 \zeta (3)+5045 \zeta (5)+3955 \zeta (7)+1890 \zeta (9)-8602)+7356 \zeta (3)+9440 \zeta (5)  \nonumber \\ &+12740 \zeta (7)+11760 \zeta (9)-33879\bigg)  \nonumber \\  B_{321} =&   -\frac{160 (89 \zeta (3)+\zeta (5)+128)}{s^3 N_f^7} \qquad \qquad B_{312}= -\frac{16 (693 \zeta (3)+115 \zeta (5)-128)}{s^4 N_c N_f^5} \nonumber \\  B_{240} =&  -\frac{64}{s^2 N_f^{10}} \bigg(4 s (3840 \zeta (3)+2240 \zeta (5)+1050 \zeta (7)+315 \zeta (9)+576) N_f+32 (30 \zeta (3)+5 \zeta (5)+12) N_f^2 \nonumber \\ &+18 \left(497 s^2+804\right) \zeta (3)+10 \left(933 s^2+1759\right) \zeta (5)+315 \left(22 s^2+69\right) \zeta (7) +3150 \left(s^2+6\right) \zeta (9) \nonumber \\ & -29977 s^2-61828\bigg)  \nonumber \\  B_{204} =&  \frac{5 (64 \zeta (5)-49 \zeta (7))}{2 s^5 N_c^3 N_f^2} \qquad \qquad B_{231} = -\frac{320 (89 \zeta (3)+\zeta (5)+128)}{s^2 N_f^8} \nonumber \\  B_{213} =& \frac{60 (64 \zeta (3)-40 \zeta (5)-7 \zeta (7))}{s^4 N_c^2 N_f^4} \qquad \qquad B_{222} = -\frac{48 (693 \zeta (3)+115 \zeta (5)-128)}{s^3 N_c N_f^6}  \nonumber \\  B_{150} =& \frac{128}{s N_f^{11}} \bigg(-2 N_f \left((32 (15 \zeta (3)+5 \zeta (5)+4)+35 \zeta (7)) N_f+4 s (480 \zeta (3)+320 \zeta (5)+175 \zeta (7)+63 \zeta (9)+64)\right) \nonumber \\ &+5148 s^2-7002 \zeta (3)-8006 \zeta (5)-9310 \zeta (7)-7812 \zeta (9)+28127\bigg) \nonumber 
\end{align}

\begin{align}
     B_{105} =& \frac{7 (10 \zeta (7)-9 \zeta (9))}{2 s^5 N_c^4 N_f} \qquad \qquad B_{141}= -\frac{320 (89 \zeta (3)+\zeta (5)+128)}{s N_f^9}  \nonumber \\  B_{114} =& \frac{640 \zeta (5)-490 \zeta (7)}{s^4 N_c^3 N_f^3} \qquad \qquad B_{132}= -\frac{64 (693 \zeta (3)+115 \zeta (5)-128)}{s^2 N_c N_f^7}  \nonumber \\  B_{123} =&  \frac{120 (64 \zeta (3)-40 \zeta (5)-7 \zeta (7))}{s^3 N_c^2 N_f^5} \qquad \qquad B_{006} = \frac{21 \zeta (9)}{4 s^5 N_c^5}  \nonumber \\  B_{060} =& -\frac{128}{N_f^{12}} \bigg(2 (320 \zeta (3)+160 \zeta (5)+70 \zeta (7)+21 \zeta (9)+64) N_f^2+2654 \zeta (3) +2882 \zeta (5)+3220 \zeta (7) \nonumber \\ &+2646 \zeta (9)-10191\bigg)    \nonumber \\  B_{051} =& -\frac{128 (89 \zeta (3)+\zeta (5)+128)}{N_f^{10}} \qquad \qquad B_{015} = \frac{70 \zeta (7)-63 \zeta (9)}{s^4 N_c^4 N_f^2}  \nonumber \\  B_{042} =& -\frac{32 (693 \zeta (3)+115 \zeta (5)-128)}{s N_c N_f^8} \qquad \qquad B_{024} = \frac{640 \zeta (5)-490 \zeta (7)}{s^3 N_c^3 N_f^4}  \nonumber \\  B_{033} =& \frac{80 (64 \zeta (3)-40 \zeta (5)-7 \zeta (7))}{s^2 N_c^2 N_f^6} \,.
\end{align}

\end{document}